\documentclass[preprint,12pt]{elsarticle}

\usepackage{amssymb}
\usepackage{amsmath}
\usepackage{fullpage}
\usepackage{booktabs}
\usepackage{rotating}
\usepackage{longtable}
\usepackage{multicol}  

\usepackage{xcolor}
\newcommand*\update[1]{\textcolor{black}{#1}}

\journal{International Journal of Heat and Mass Transfer}

\begin{document}

\begin{frontmatter}



\title{Modelling laminar flow in V-shaped filters integrated with catalyst technologies for atmospheric pollutant removal}

\author[label1,label2,label3]{Samuel D. Tomlinson}
\author[label1,label2]{Aliki M. Tsopelakou}
\author[label1,label2]{Tzia Ming Onn}
\author[label1]{Steven R. H. Barrett}
\author[label1,label2,label4]{Adam M. Boies}
\author[label1,label2]{Shaun Fitzgerald}
\affiliation[label1]{organization={Department of Engineering, University of Cambridge},
            city={Cambridge},
            postcode={CB2 1PZ},
            country={UK}}
\affiliation[label2]{organization={Centre for Climate Repair},
            city={Cambridge},
            postcode={CB3 0WA},
            country={UK}}
\affiliation[label3]{organization={School of Computing and Mathematical Sciences, University of Greenwich},
            city={London},
            postcode={SE10 9LS},
            country={UK}}
\affiliation[label4]{organization={Department of Mechanical Engineering, Stanford University},
            city={Stanford},
            postcode={CA 94305},
            country={USA}}

\begin{abstract}

Atmospheric pollution from particulate matter, volatile organic compounds and greenhouse gases is a critical environmental and public health issue, leading to respiratory diseases and climate change. 
A potential mitigation strategy involves utilising ventilation systems, which process large volumes of indoor and outdoor air and remove particulate pollutants through filtration.
However, the integration of catalytic technologies with filters in ventilation systems remains underexplored, despite their potential to simultaneously remove particulate matter and gases, as seen in flue gas treatment and automotive exhaust systems.
In this study, we develop a predictive, long-wave model for V-shaped filters, with and without separators.
The model, validated against experimental and numerical data, provides a framework for enhancing flow rates by increasing fibre diameter and porosity while reducing aspect ratio and filter thickness. 
These changes lead to increased permeability, which lowers energy requirements. 
However, they also reduce the pollutant removal efficiency, highlighting the trade-off between flow, filtration performance and operational costs.
Leveraging the long-wave model alongside experimental results, we estimate the maximum potential removal rate ($4.5\times10^{-3}$ GtPM$_{2.5}$, $6.4\times10^{-3}$ GtNO$_{\text{x}}$, $2.0\times10^{-2}$ GtCH$_{4}$ per year; $1.6\times10^{0}$ GtCO$_{2}$e per year, 20-year GWP for CH$_4$) and minimum cost (\$$3.4\times10^{3}$ per tNO$_{\text{x}}$, \$$1.1\times10^{3}$ per tCH$_{4}$; \$$1.3\times10^{1}$ per tCO$_{2}$e) if a billion V-shaped filters integrated with catalytic enhancements were deployed in operation. 
These findings highlight the feasibility of catalytic filters as a scalable, high-efficiency solution for improving air quality and mitigating atmospheric pollution.

\end{abstract}







\end{frontmatter}


%

\begin{table}[t!]
\scriptsize
\centering
\begin{tabular}{@{}p{2cm}@{}p{5.3cm}@{}p{2cm}@{}p{5.3cm}@{}}

\toprule
\multicolumn{4}{l}{\textbf{Nomenclature}} \\[0.5em]

\multicolumn{2}{l}{\textbf{Roman letters}} & \multicolumn{2}{l}{\textbf{Greek letters continued}} \\

\hspace{.12cm} $\hat{A}$ & Cross-sectional area [m$^2$] & \hspace{.12cm} $\epsilon$ & Aspect ratio [--] \\
\hspace{.12cm} $\hat{c}$ & Pollutant concentration [mol/m$^3$] & \hspace{.12cm} $\eta$ & Removal efficiency [--] \\
\hspace{.12cm} $\hat{d}$ & Fibre diameter [m] & \hspace{.12cm} $\hat{\rho}$ & Fluid density [kg/m$^3$] \\
\hspace{.12cm} $\hat{H}$ & Half-period height [m] & \hspace{.12cm} $\phi$ & Porosity [--] \\
\hspace{.12cm} $\hat{k}$ & Permeability of the filter sheet [m$^2$] & \multicolumn{2}{l}{\textbf{Subscripts}} \\
\hspace{.12cm} $\hat{L}$ & Half-period length [m] & \hspace{.12cm} in & Inlet \\
\hspace{.12cm} $N$ & Number of half-periods [--] & \hspace{.12cm} out & Outlet \\
\hspace{.12cm} $\hat{p}$ & Pressure field [Pa] & \hspace{.12cm} 0 & Leading-order approximation \\
\hspace{.12cm} $\hat{Q}$ & Volumetric flow rate [m$^3$/s] & \hspace{.12cm} $x,\,y,\,z$ & Partial derivatives in respective directions \\
\hspace{.12cm} $\hat{q}$ & Pollutant removal rate [mol/s] & \multicolumn{2}{l}{\textbf{Abbreviations}} \\
\hspace{.12cm} $\hat{s}$ & Filter sheet location [m] & \hspace{.12cm} PM$_\text{2.5}$ & Particulate matter $< 2.5~\mu$m \\
\hspace{.12cm} $\hat{t}$ & Filter thickness [m] & \hspace{.12cm} NO$_\text{x}$ & Nitrogen oxides \\
\hspace{.12cm} $\hat{u},\,\hat{v},\,\hat{w}$ & Velocity components (streamwise, normal, spanwise) [m/s] & \hspace{.12cm} CH$_4$ & Methane \\
\hspace{.12cm} $\hat{W}$ & Half-period width [m] & \hspace{.12cm} GHGs & Greenhouse gases \\
\hspace{.12cm} $\hat{x},\,\hat{y},\,\hat{z}$ & Cartesian coordinates [m] & \hspace{.12cm} VOCs & Volatile organic compounds \\

\multicolumn{2}{l}{\textbf{Greek letters}} & \hspace{.12cm} CFD & Computational fluid dynamics \\
\hspace{.12cm} $\hat{\beta}$ & Inertial resistance coefficient [s$^2$/m$^2$] & \hspace{.12cm} MERV & Minimum efficiency reporting value \\
\hspace{.12cm} $\hat{\kappa}$ & Permeance [m/(Pa$\cdot$s)] & \hspace{.12cm} SEM & Scanning electron microscopy \\
\hspace{.12cm} $\hat{\mu}$ & Dynamic viscosity [Pa$\cdot$s] & \hspace{.12cm} HEPA & High-efficiency particulate air \\
\bottomrule
\end{tabular}
\end{table}

\section{Introduction}

Atmospheric pollution poses a severe threat to human health, ecosystems and climate stability, making its mitigation a global priority~\cite{kumar2021climate, jones2023national, manisalidis2020environmental, world2021global}.
Particulate matter (PM), volatile organic compounds (VOCs) and nitrogen oxides (NO$_\text{x}$) contribute to respiratory and cardiovascular diseases, premature mortality and environmental degradation~\cite{sosa2017human, world2021global, polichetti2009effects}.
In parallel, the accumulation of GHGs such as carbon dioxide (CO$_2$), methane (CH$_4$) and nitrous oxide (N$_2$O) is accelerating global warming and climate instability~\cite{prentice2001carbon, kumar2021climate}.
While strategies to mitigate outdoor pollution are essential, it is also important to manage indoor air quality, given that people now spend the majority of their time indoors.

Ventilation systems play a pivotal role in indoor air quality by filtering airborne pollutants and reducing exposure to harmful pathogens~\cite{tran2020indoor, elsaid2021indoor, dimitroulopoulou2023indoor}.
Ubiquitous in residential, commercial and industrial environments, ventilation systems process large volumes of indoor and outdoor air~\cite{tomlinson2025harnessing}, predominantly removing PM and typically leaving GHGs and VOCs unaddressed~\cite{del2002air, brincat2016review, liu2020progress}.
Similar air-processing systems are found in applications such as mines, combustion-intake systems, cooling towers and agricultural facilities, where they help manage air quality by moving PM and gases such as CH$_4$.

Catalytic filters are used in flue gas treatment to remove NO$_\text{x}$ and dust, thereby reducing the operational costs associated with pollution control~\cite{li2022recent}. 
A similar approach is employed in automotive exhaust systems, where catalytic diesel particulate filters capture PM while simultaneously reducing NO$_\text{x}$ emissions through selective catalytic reduction~\cite{van2001science}.
\update{Advances in catalytic and nanostructured materials has accelerated the development of multifunctional air-purification systems capable of removing pathogens, particulates and gases ~\cite{malayeri2019modeling, attia2024nano, shin2025photo, songxuan2025application, wang2025nanofibrillated}. 
Specifically, Malayeri \textit{et al.} ~\cite{malayeri2019modeling} review photocatalytic VOC degradation; Attia \textit{et al.}~\cite{attia2024nano} present nano-carbon filters for particulate removal; Shin \textit{et al.}~\cite{shin2025photo} report photo-regenerable antimicrobial filters; Songxuan \textit{et al.}~\cite{songxuan2025application} demonstrate pilot-scale catalytic multi-pollutant removal; and Wang \textit{et al.}~\cite{wang2025nanofibrillated} develop nanofiber networks for enhanced pollutant capture. 
}
\update{Regulatory bodies are taking note: for example, the California Air Resources Board is funding research on advanced filters for non‑PM pollutants, reflecting the growing interest in translating these laboratory-scale advances into real-world applications~\cite{CARB2025_nonPMfilters}.}
Motivated by these successes, recent studies have proposed deploying catalytic filters that combine PM capture with catalytic conversion for GHGs or VOCs in ventilation systems~\cite{tomlinson2025harnessing}.
\update{Integrating catalytic filtration within ventilation systems provides a potentially scalable route for air-quality improvement and GHG mitigation.}
However, the performance and design of such catalytic filtration systems remain under-explored, particularly in the context of flow optimisation and and filter design.
Another similar integration of removal technologies was proposed during the COVID pandemic, where a number of approaches to tackle virus elimination were proposed such as high grade filters, UV-C light and antimicrobial coatings~\cite{thornton2022impact, watson2022efficacy}.
Furthermore, filters are, or should be, replaced at regular intervals, which facilitates the integration of new technologies into existing systems. 
As a result, the design life of a new product can be shorter and more readily achievable.

\begin{figure*}[t!]
    \centering
    \includegraphics[width=\linewidth]{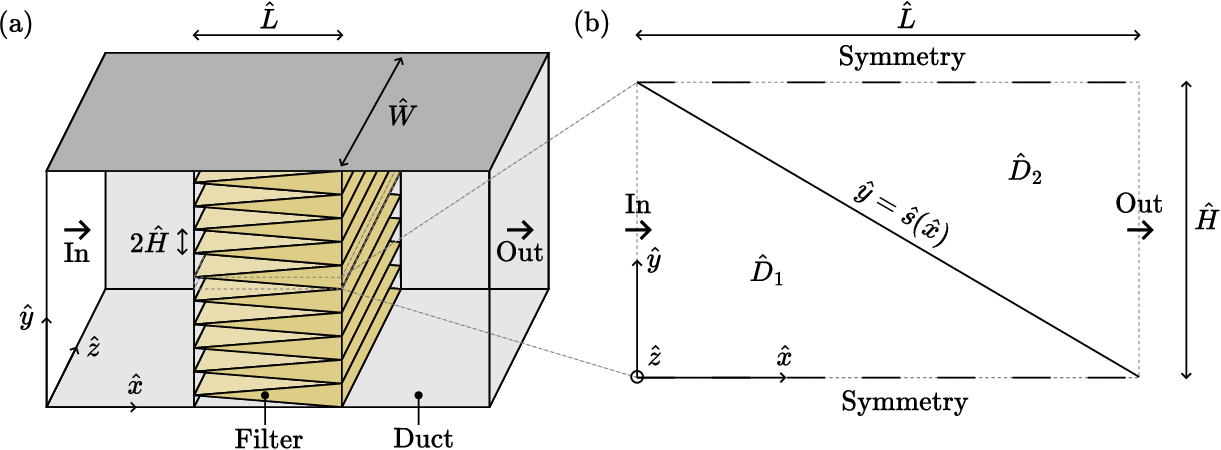} \\
    \includegraphics[width=\linewidth]{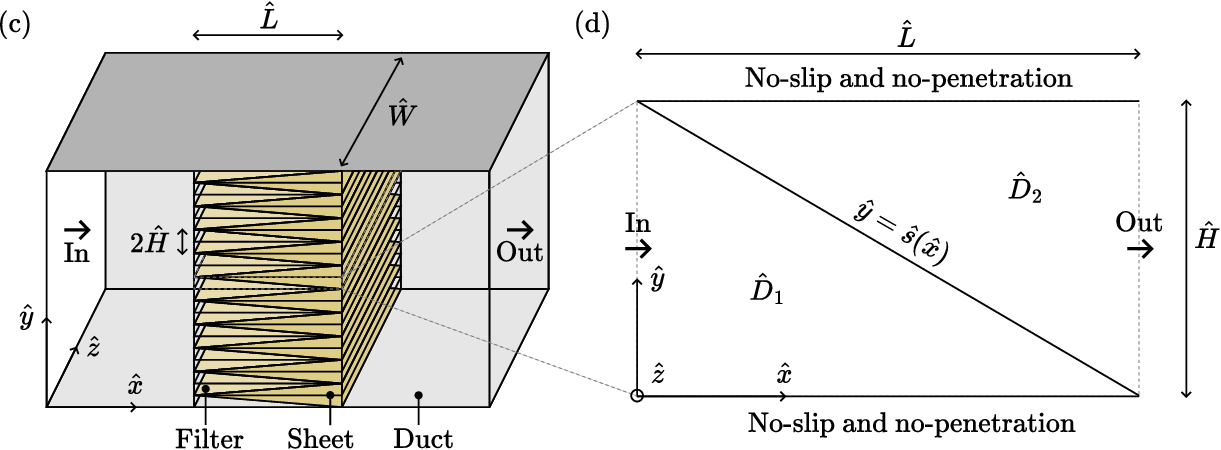} \\
    \small \hspace{.5cm} (e) \hfill \hfill \hfill\\
    \includegraphics[width=.86\linewidth]{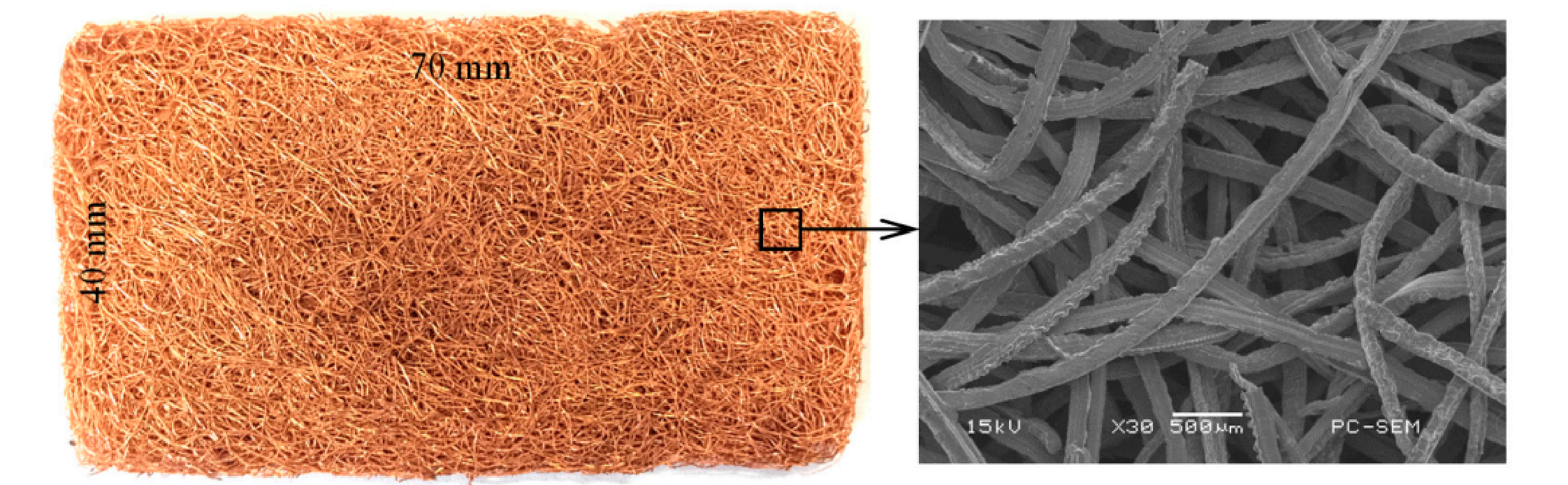}
    \caption{
    A pressure-driven airflow through a (a) duct containing a V-shaped filter, (b) a single half period of the V-shaped filter, (c) duct containing a separated V-shaped filter and (d) a single half period of the separated V-shaped filter.
    In (b, d), we position the origin of the coordinate system, $(\hat{x}, \, \hat{y}, \, \hat{z})$, at the start of the V-shaped filter, midway between the filter sheets and transverse width. 
    Each period of the V-shaped filter has length $\hat{L}$, height 2$\hat{H}$ and width $\hat{W}$.
    The location of the filter sheet is given by $\hat{y} = \hat{s}(\hat{x})$; the area below and above the filter sheet defines subdomains $\hat{D}_1$ and $\hat{D}_2$, respectively.
    (e) Photograph of a $\phi = 80$\% porosity fibre sheet and SEM image of the microstructure with fibres of around $\hat{d} = 100 \mu$m thickness, reproduced from Huang \textit{et al.}~\cite{huang2021effective}, licensed under CC BY 4.0.
    }
    \label{fig:not_seperated}
\end{figure*}

Filtration efficiency is governed by fluid-structure interactions within a porous medium, with filter sheets classified as either fibre-based or pore-based (see Fig.~\ref{fig:not_seperated}). 
While fibre-based filters are easier to manufacture, they offer less control and can have a higher environmental impact than pore-based alternatives, as they are more prone to clogging and may often be discarded. 
In contrast, pore-based filters can be cleaned and reused, reducing their environmental impact~\cite{sutherland2011filters, pereira2021optimising}.
Pleating the filter sheet increases the surface area, thereby enhancing airflow or reducing pressure drop~\cite{sutherland2011filters, mrad2021local}, but may introduce velocity gradients across the filter, leading to variations in pollutant removal efficiency and clogging.
Predicting the permeability of porous media is essential for optimising filter performance, with the Kozeny-Carman relation providing an empirical link between permeability, porosity and structural properties of the filter sheet~\cite{kozeny1927ueber, carman1937fluid}.
The theory of filtration relates permeability to particle collection on isolated fibres and identifies the main capture mechanisms in fibrous media as interception, diffusion and inertial impaction~\cite{hinds2022aerosol}.
As particles are filtered out, the filter sheet blocks, either through the accumulation of particles on the fibres or within the pores, which affects the permeability and hence the airflow rate through the sheet~\cite{teng2022research, zhang2022operating, lowther2023factors}.
The level of filtration can be measured in terms of the removal efficiency, which represents the percentage of particles removed (e.g., 99.97\%) larger than a certain size (e.g., $>$0.3$\mu$m)~\cite{zhang2022operating}.

Advances in filtration have combined modelling, computational simulations and experimental validation.
Prior studies have resolved the velocity and pressure field in various V- and U-shaped filter geometries in laminar and turbulent flows, using reduced models and computational fluid dynamics (CFD) simulations~\cite{del2002air, tronville2003minimization, rebai2010semi, herterich2017optimizing, theron2017numerical, wang2017role, xu2017pressure, mrad2021local, teng2022research, zhang2022operating}.
Pereira \textit{et al.}~\cite{pereira2021optimising} refined these modelling approaches using a laminar linear long-wave theory within a periodic filter with dead ends, identifying designs to maximise airflow rates through the device.
While CFD simulations enhance the understanding of airflow through filters, their high cost limits their use for optimisation.
Experimental investigations using techniques like Hot Wire Anemometry and Particle Image Velocimetry~\cite{del2002air, 
rebai2010semi, al2011effect, feng2014assessment, youssef2016experimental, theron2017numerical, kang2020characterization, mrad2021local, teng2022research, zhang2022operating} have validated these models across various filter designs.

\update{This study focuses on laminar experimental datasets for V-shaped filters reported in the literature, providing measurements of pressure drops and flow fields under conditions where the effects of dust loading are negligible.} 
A brief summary follows.
Fabbro \textit{et al.}~\cite{del2002air} explored the relationship between pressure drop and velocity through a V-shaped filter across filter geometries and efficiencies. 
Mrad \textit{et al.}~\cite{mrad2021local} examined the streamwise velocity field upstream and downstream of the V-shaped filter, comparing their results to simulations performed for laminar and turbulent flows.
Zhang \textit{et al.}~\cite{zhang2022operating} investigated the airflow rate of V-shaped and cylindrical filters, measuring the deposition of particles in filters with different porosities and fibre diameters.
Al-Attar \textit{et al.}~\cite{al2011effect} investigated the effects of pleat density, pleat orientation and face velocity on V-shaped filters from ventilation and gas-turbine applications.

Despite these advancements, uncertainties remain in filter performance.
Firstly, existing models are validated on a case-by-case basis rather than being predictive across devices to allow for the optimisation of filter design.
Secondly, a balance must be struck between high-resolution simulations, which can be computationally expensive, and reduced-order models, which may lack precision.
Thirdly, the influence of separators, which structurally support filters and impact airflow dynamics, remains unquantified.
Lastly, the integration of catalytic technologies into filters will introduce modifications to the filter-sheet structure and pollutant removal efficiency, as discussed in~\cite{songxuan2025application}.

To address these uncertainties, we develop a predictive non-linear long-wave model for laminar flow in V-shaped filters, incorporating the effects of separators.
The long-wave model incorporates relations that enable the flow rate and permeability of the filter to be determined from measurable design parameters: fibre diameter, porosity, filter length, height and thickness.
We validate the long-wave model using numerical and experimental data~\cite{del2002air, al2011effect, mrad2021local, pereira2021optimising, zhang2022operating}, demonstrating its applicability across filter designs.
\update{Building on the long-wave analysis of Pereira \textit{et al.}~\cite{pereira2021optimising}, this model provides closed-form solutions that account for separators and weak-inertia, unifies the Kozeny--Carman permeability relation with an empirical efficiency model, enabling application-oriented predictions of trade-offs between permeability, energy consumption and pollutant removal. 
We also examine how design parameters influence the potential removal efficiency of PM and different gases for V-shaped filters integrated with catalytic technologies. 
}
Our findings suggest that catalytic filters could serve as a solution for improving indoor and outdoor air quality, with implications for public health and climate mitigation.

\section{Methods} \label{sec:formulation}

In this study, we analyse airflow through a V-shaped filter, where the geometry and fibre-sheet properties will influence the velocity and pressure field (see Fig. \ref{fig:not_seperated}a, c).
Table \ref{tab:characteristic_values} summarises the key flow characteristics for V-shaped filters, including length, velocity, pressure and material properties used in the analysis.
The streamwise, wall-normal and spanwise directions are represented by $\hat{x}$-, $\hat{y}$- and $\hat{z}$-coordinates, where hats indicate dimensional quantities. 
Assuming the fluid is incompressible and Newtonian, consistent with standard models, the velocity field is given by $\boldsymbol{\hat{u}} = (\hat{u}(\hat{x}, \, \hat{y}, \, \hat{z}), \, \hat{v}(\hat{x}, \, \hat{y}, \, \hat{z}), \, \hat{w}(\hat{x}, \, \hat{y}, \, \hat{z}))$ and the pressure field is $\hat{p} = \hat{p}(\hat{x}, \, \hat{y}, \, \hat{z})$. 
The V-shaped filter is assumed to be periodic in the normal direction and symmetric around the centre line of each period.
Consequently, we focus on a single half-period of dimensions $\hat{L}$ in length, $2\hat{H}$ in height and $\hat{W}$ in width.
Assuming the filter sheet thickness ($\hat{t}$) is negligible relative to the other length scales of the filter (Table \ref{tab:characteristic_values}), its location, $\hat{y} = \hat{s}(\hat{x})$, is given by
\begin{equation} \label{eq:dim_sheet}
    \hat{s}(\hat{x}) = \hat{H}(1 - \hat{x}/\hat{L}).
\end{equation}
To account for the different flow characteristics above and below the filter sheet, we divide the half-period into two subdomains
\begin{subequations} \label{eq:dim_dom_1}
\begin{align}
    \hat{D}_1 &= \{\hat{x}\in [0,\,\hat{L}]\}\times\{\hat{y}\in[0,\,\hat{s}]\}\times\{\hat{z}\in [0,\,\hat{W}]\}, \\
    \hat{D}_2 &= \{\hat{x}\in [0,\,\hat{L}]\}\times\{\hat{y}\in[\hat{s},\,\hat{H}]\}\times\{\hat{z}\in [0,\,\hat{W}]\},
\end{align}
\end{subequations}
as in Fig. \ref{fig:not_seperated}(b, d).

\begin{table}[t!]
\centering
\renewcommand{\arraystretch}{1.3} 
\setlength{\tabcolsep}{4pt} 
\scriptsize
\begin{tabular}{lccc}
    \toprule
    Name & Symbol & Unit & Value \\
    Pressure difference & $\Delta \hat{P}$ & Pa & 300 \\
    Half-period length & $\hat{L}$ & m & 0.1 \\
    Half-period height & $\hat{H}$ & m & 0.004 \\
    Half-period width & $\hat{W}$ & m & 0.1 \\
    Thickness & $\hat{t}$ & m & $10^{-3}$ \\
    Permeability & $\hat{k}$ & m$^2$ & $10^{-9}$ \\
    Number of half-periods & $N$ & -- & 200 \\
    \bottomrule
    \end{tabular}
    \caption{A summary of characteristic pressure, length and material scales characteristic of V-shaped filters~\cite{del2002air, al2011effect, mrad2021local, pereira2021optimising, zhang2022operating,tomlinson2025harnessing}.}
    \label{tab:characteristic_values}
\end{table}

The flow enters the half period through $\hat{D}_1$ at $\hat{x} = 0$ with a constant inlet pressure 
\begin{equation} \label{eq:dim_pres_in}
    \hat{p} = \hat{p}_\text{in},
\end{equation} 
and exits through $\hat{D}_2$ at $\hat{x} = \hat{L}$ with a constant outlet pressure 
\begin{equation} \label{eq:dim_pres_out}
    \hat{p} = \hat{p}_\text{out}.
\end{equation} 
The imposed pressure drop, $\Delta \hat{p} \equiv \hat{p}_\text{in} - \hat{p}_\text{out} > 0$, drives the airflow.
To capture airflow behaviour, we solve the Navier–Stokes equations in $\hat{D}_1$ and $\hat{D}_2$
\begin{subequations} \label{eq:dim_mass}
\begin{align}
    \hat{u}_{\hat{x}} + \hat{v}_{\hat{y}} + \hat{w}_{\hat{z}} &= 0, \\
    \hat{\mu}(\hat{u}_{\hat{x}\hat{x}} + \hat{u}_{\hat{y}\hat{y}} + \hat{u}_{\hat{z}\hat{z}}) - \hat{p}_{\hat{x}} &= \hat{\rho}(\hat{\boldsymbol{u}} \cdot \hat{\boldsymbol{\nabla}})\hat{u}, \\
    \hat{\mu}(\hat{v}_{\hat{x}\hat{x}} + \hat{v}_{\hat{y}\hat{y}} + \hat{v}_{\hat{z}\hat{z}}) - \hat{p}_{\hat{y}} &= \hat{\rho}(\hat{\boldsymbol{u}} \cdot \hat{\boldsymbol{\nabla}})\hat{v}, \\
    \hat{\mu}(\hat{w}_{\hat{x}\hat{x}} + \hat{w}_{\hat{y}\hat{y}} + \hat{w}_{\hat{z}\hat{z}}) - \hat{p}_{\hat{z}} &= \hat{\rho}(\hat{\boldsymbol{u}}\cdot \hat{\boldsymbol{\nabla}})\hat{w},
\end{align}
\end{subequations}
where $\hat{\mu}$ is the dynamic viscosity and $\hat{\rho}$ is the density of air.
Since filters are usually enclosed within rigid housings, at the ends of the V-shaped filter in the transverse direction, we impose no-slip and no-penetration conditions at $\hat{z} = 0, \, \hat{W}$:
\begin{equation} \label{eq:dim_end}
    \hat{u}  = \hat{v} = \hat{w} = 0.
\end{equation}
We examine two common filter configurations (Fig. \ref{fig:not_seperated}a--d), one with and one without separators, allowing for the comparison of flow behaviour from small-scale residential to large-scale industrial systems. 
The configuration affects the imposed boundary conditions and therefore the airflow behaviour.
If the V-shaped filter does not have separators (Fig. \ref{fig:not_seperated}a--b), we impose symmetry conditions across each half-period at $\hat{y} = 0, \, \hat{H}$:
\begin{equation} \label{eq:dim_per}
    \hat{u}_{\hat{y}}  = \hat{v} = \hat{w}_{\hat{y}} = 0.
\end{equation}
If the V-shaped filter has separators (Fig. \ref{fig:not_seperated}c--d), we impose no-slip and no-penetration conditions at $\hat{y} = 0, \, \hat{H}$:
\begin{equation} \label{eq:dim_noslip}
    \hat{u}  = \hat{v} = \hat{w} = 0,
\end{equation}
To account for both viscous and inertial resistance effects introduced by the filter sheet, we employ the Darcy–Forchheimer model~\cite{ingham2005transport}. 
\update{Typical inertial resistance coefficients for fibrous materials are, for example, $\hat{\beta}=10^{-5}$ to $10^{-3}$ s/m, based on Tamayol \textit{et al.}~\cite{tamayol2012effects}. 
Based on Shin~\cite{shin2022application}, when the Reynolds number is calculated using the apparent flow velocity (0.01 to 0.1 m/s) from Figure 2, the thickness (0.5 mm) and the height of the material (4 mm), the Darcy term dominates for approximately $Re = \hat{\rho}\hat{U}\hat{H}/\hat{\mu} <50$; inertial effects become comparable to viscous forces for around $50<Re<500$; and nonlinear pressure--flow behaviour arises due to the Forchheimer term when approximately $Re>500$. 
}
This model is widely used in porous-media analysis, incorporating permeability ($\hat{k}$) effects via the permeance ($\hat{\kappa} = \hat{k}/(\hat{\mu}\hat{t})$) and pressure losses at higher flow velocities through the inertial resistance ($\hat{\beta}$). 
At $\hat{y} = \hat{s}(\hat{x})$, we have
\begin{equation} \label{eq:dim_sheet_1}
    \hat{\boldsymbol{u}}\cdot \hat{\boldsymbol{n}} = \hat{\kappa}(\hat{p}^- - \hat{p}^+) - \hat{\beta}(\hat{u}^2 + \hat{v}^2 + \hat{w}^2)^{1/2}\hat{\boldsymbol{u}}\cdot \hat{\boldsymbol{n}},
\end{equation}
where $\hat{\boldsymbol{n}} \equiv (\hat{H}/\hat{L}, \, 1, \, 0)$ is the normal vector to the filter sheet and the superscript $-$ (+) indicates evaluation in $\hat{D}_1$ ($\hat{D}_2$).
For simplicity, we impose no tangential velocity at $\hat{y} = \hat{s}(\hat{x})$
\begin{equation} \label{eq:dim_sheet_2_a}
    \hat{\boldsymbol{u}}\cdot \hat{\boldsymbol{t}}_x = \hat{\boldsymbol{u}}\cdot \hat{\boldsymbol{t}}_z = 0,
\end{equation}
where $\hat{\boldsymbol{t}}_x \equiv ( 1, \, -\hat{H}/\hat{L}, \, 0)$ and $\hat{\boldsymbol{t}}_z \equiv (0, \, 0, \, 1)$ are the tangent vectors to the filter sheet. 
The flow rate through the filter half-period is constant as mass is conserved and is given by 
\begin{equation} \label{eq:dim_flux}
    \hat{Q} = \int_{\hat{y} = 0}^{\hat{H}} \int_{\hat{z} = 0}^{\hat{W}} \hat{u}(0, \, \hat{y},\, \hat{z}) \, \text{d} \hat{y} \text{d} \hat{z}.
\end{equation}
Therefore, assuming that the inlet pollutant concentration $\hat{c}$ is uniform in $\hat{D}_1$, as there are no sources or sinks other than the filter sheet, the potential pollutant removal rate is 
\begin{equation}
    \hat{q} = \eta \hat{Q} \hat{c},
\end{equation}
where $\hat{Q} \hat{c}$ is the flow rate of pollutant to the filter sheet and $\eta$ is the removal efficiency of the pollutant-removal technology, defined as the fraction of pollutant molecules removed upon contact with the filter sheet.
We aim to determine the dependence of $\hat{Q}$ and $\hat{q}$ on the dimensional parameters of the problem, $\hat{L}$, $\hat{H}$, $\hat{W}$, $\Delta\hat{p}$, $\hat{\mu}$, $\hat{\kappa}$, $\hat{\rho}$, $\hat{\beta}$, $\eta$ and $\hat{c}$, in V-shaped filter configurations.

\subsection{Non-dimensionalisation} 

We begin by non-dimensionalising the governing equations \eqref{eq:dim_sheet}--\eqref{eq:dim_flux} using the characteristic length ($\hat{L}$), velocity ($\hat{U}$) and pressure ($\epsilon^2 \hat{L}/\hat{\mu}\hat{U}$) scales of the flow:
\begin{equation} \label{eq:non_dim}
    x = \frac{\hat{x}}{\hat{L}}, \quad y = \frac{\hat{y}}{\epsilon \hat{L}}, \quad z = \frac{\hat{z}}{\hat{L}}, \quad u = \frac{\hat{u}}{\hat{U}}, \quad v = \frac{\hat{v}}{\epsilon \hat{U}}, \quad w = \frac{\hat{w}}{\hat{U}}, \quad p = \frac{\epsilon^2 \hat{L} (\hat{p} - \hat{p}_{\text{out}})}{\hat{\mu}\hat{U}},
\end{equation}
where $\epsilon \equiv \hat{H} / \hat{L}$ is the aspect ratio of the half period.
The location of the filter sheet \eqref{eq:dim_sheet} becomes
\begin{equation} \label{eq:sheet}
    s(x) = 1 - x.
\end{equation}
The subdomains \eqref{eq:dim_dom_1} become
\begin{subequations} \label{eq:dom_1}
\begin{align}
    D_1 &= \{x\in [0,\,1]\}\times\{y\in[0,\,s]\}\times\{z\in [0, \, W]\}, \\
    D_2 &= \{x\in [0,\,1]\}\times\{y\in[s,\,1]\}\times\{z\in [0, \, W]\},
\end{align}
\end{subequations}
where $W = \hat{W}/\hat{L}$ is the non-dimensional width of the filter. 
At the inlet, $x=0$, the pressure condition \eqref{eq:dim_pres_in} gives
\begin{equation} \label{eq:in}
    p = 1,
\end{equation} 
and at the outlet, $x=1$, the pressure condition \eqref{eq:dim_pres_out} gives
\begin{equation} \label{eq:out}
    p = 0.
\end{equation} 
In $D_1$ and $D_2$, the bulk equations \eqref{eq:dim_mass} become 
\begin{subequations} \label{eq:mass}
\begin{align}
    u_x + v_y + w_z &= 0, \\
    \epsilon^2 (u_{xx} + u_{zz}) + u_{yy} - p_x &= \epsilon^2 {Re}(\boldsymbol{u} \cdot \boldsymbol{\nabla})u, \\
    \epsilon^4 (v_{xx} + v_{zz}) + \epsilon^2 v_{yy} - p_y &= \epsilon^4 {Re}(\boldsymbol{u} \cdot \boldsymbol{\nabla})v, \\
    \epsilon^2 (w_{xx} + w_{zz}) + w_{yy} - p_z &= \epsilon^2 {Re}(\boldsymbol{u} \cdot \boldsymbol{\nabla})w,
\end{align}
\end{subequations}
where ${Re} = \hat{\rho} \hat{U} \hat{H} / \hat{\mu}$ is the Reynolds number. 
At $z = 0, \, W$, the end conditions \eqref{eq:dim_end} give
\begin{equation} \label{eq:end}
    u = v = w = 0.
\end{equation}
At $y=0,\,1$, the symmetry conditions \eqref{eq:dim_per} become
\begin{equation} \label{eq:per}
    u_y  = v = w_y = 0,
\end{equation}
and the no-slip and no-penetration conditions \eqref{eq:dim_noslip} become
\begin{equation} \label{eq:noslip}
    u = v = w = 0.
\end{equation}
At $y = s(x)$, the normal flow condition \eqref{eq:dim_sheet_1} gives
\begin{equation} \label{eq:sheet_1}     \boldsymbol{u}\cdot\boldsymbol{n} = \kappa (p^- - p^+) - \beta (u^2 + \epsilon^2 v^2 + w^2)^{1/2}\boldsymbol{u}\cdot\boldsymbol{n},
\end{equation}
where $\boldsymbol{n} = (1,\, 1,\, 0)$ is the unit normal vector at the filter-sheet surface, $\kappa = \hat{\mu}\hat{\kappa}/\epsilon^3 \hat{L}$ is the permeance, $k = \kappa / (\hat{t}/\hat{L})$ is the permeability and $\beta= \hat{\beta} \hat{U}$ is the inertial resistance of the filter sheet.
Parameters $\kappa$ and $k$ quantify the filter's resistance to flow through the filter sheet, and $\beta$ accounts for non-Darcian effects due to inertia at higher velocities.
The tangential flow conditions \eqref{eq:dim_sheet_2_a} give
\begin{equation} \label{eq:sheet_2_a}
    \boldsymbol{u}\cdot\boldsymbol{t}_x = \boldsymbol{u}\cdot\boldsymbol{t}_z = 0,
\end{equation}
where $\boldsymbol{t}_x \equiv (1, \, -\epsilon^2, \, 0)$ and $\boldsymbol{t}_z \equiv (0, \, 0, \, 1)$ are tangential vectors at the filter-sheet surface. 
The volumetric flow rate through the half period \eqref{eq:dim_flux} becomes
\begin{equation} \label{eq:flux}
    Q = \int_{y = 0}^{1} \int_{z = 0}^{W} u \, \text{d} y \text{d} z,
\end{equation}
and the potential removal rate of the pollutant is given by
\begin{equation} \label{eq:pollutant_removal_rate}
    q = \eta Q.
\end{equation}
In the limit $\epsilon^2 \ll 1$ and $\epsilon^2 {Re} \ll 1$, the governing equations simplify, and we proceed to solve the leading-order boundary value problem defined in \eqref{eq:in}--\eqref{eq:flux} using asymptotic and numerical methods. 
We evaluate $Q$ and $q$ through the V-shaped filter, \eqref{eq:flux}--\eqref{eq:pollutant_removal_rate}, and investigate its dependence on the dimensionless groups $\epsilon$, $W$, $\kappa$, $\beta$ and $\eta$ that characterise the geometry, filter and flow. 

\subsection{The long-wave limit}

We begin by taking the limit as the aspect ratio of the half period $\epsilon \equiv \hat{H}/\hat{L} \rightarrow 0$, substituting the expansions 
\begin{equation} \label{eq:expansion}
	(u,\,v, \,p) = (u_0,\,v_0, \,p_0) + \epsilon^2 (u_1,\,v_1, \,p_1) + ...,
\end{equation}
into the governing equations \eqref{eq:in}--\eqref{eq:sheet_2_a} and taking the leading-order approximation.
In this study, we consider the limit where $B \equiv \beta\epsilon=O(1)$ to capture the inertial effects observed in the experiments~\cite{del2002air, mrad2021local, zhang2022operating, al2011effect} that influence the flow through the filter sheet.
At the inlet, $x=0$, the pressure condition \eqref{eq:in} gives
\begin{equation} \label{eq:in_0}
    p_0 = 1,
\end{equation} 
and at the outlet, $x=1$, the pressure condition \eqref{eq:out} gives
\begin{equation} \label{eq:out_0}
    p_0 = 0.
\end{equation} 
In $D_1$ and $D_2$, the bulk equations \eqref{eq:mass} become
\begin{subequations} \label{eq:mass_0}
\begin{align}
    u_{0x} + v_{0y} + w_{0z} &= 0, \\
    u_{0yy} - p_{0x} &= 0, \\
    p_{0y} &= 0, \\
    w_{0yy} - p_{0z} &= 0.
\end{align}
\end{subequations}
The long-wave limit, $\epsilon \rightarrow 0$, leads to short Navier-Stokes regions, which ensure that the end conditions \eqref{eq:end} are satisfied.
At $y=0,\,1$, the symmetry conditions \eqref{eq:per} become
\begin{equation} \label{eq:per_0}
    u_{0y}  = v_0 = w_{0y} = 0,
\end{equation}
and the no-slip and no-penetration conditions \eqref{eq:noslip} become
\begin{equation} \label{eq:noslip_0}
    u_0 = v_0 = w_0 = 0.
\end{equation}
At $y = s(x)$, the normal flow condition \eqref{eq:sheet_1} gives
\begin{equation} \label{eq:sheet_1_0}
    v_0 = \kappa (p_0^- - p_0^+) - B v_0^2,
\end{equation}   
and the tangential flow conditions \eqref{eq:sheet_2_a} give
\begin{equation} \label{eq:sheet_2_0}
    u_0 = w_0 = 0.
\end{equation}
The flow rate through the half period \eqref{eq:flux} becomes
\begin{equation} \label{eq:flux_0}
    Q \approx \int_{y = 0}^{1} \int_{z = 0}^{W} u_0 \, \text{d} y \, \text{d} z,
\end{equation}
and the potential removal rate of pollutant remains as is given by \eqref{eq:pollutant_removal_rate}.

The transverse flow problem is not forced directly, and we expect it to relax to a solution for which $p_0 = p_0(x)$ and $w_0 = 0$.
First, we consider the V-shaped filter without separators.
We can solve \eqref{eq:mass_0} subject to \eqref{eq:per_0}--\eqref{eq:sheet_2_0} in $D_1$ to give
\begin{subequations} \label{eq:v_per_1}
\begin{align}
    u_0 &= p_{0x}(y^2 - s^2)/2, \\
    v_0 &=  - p_{0xx} y^3/6 + (p_{0xx}s^2/2 + p_{0x} s s_x)y, 
\end{align}
\end{subequations} 
and in $D_2$ to give
\begin{subequations} \label{eq:v_per_2}
\begin{align}
    u_0 &= p_{0x}((y-1)^2 - (s-1)^2)/2, \\
    v_0 &= (-p_{0xx}((y-1)^2/3 - (s-1)^2) + 2 p_{0x} (s-1) s_x)(y-1)/2.
\end{align}
\end{subequations}
Substituting the velocity field \eqref{eq:v_per_1}--\eqref{eq:v_per_2} into \eqref{eq:sheet_1_0} gives a system of ordinary differential equations (ODEs) along $y = s(x)$.
We can solve this system of ODEs to give the solution in $D_1$ and $D_2$ as $p_0 = p_0(x)$.
Along $y = s(x)$, 
\begin{subequations} \label{eq:bvp_1}
\begin{align}
   \mathcal{L}_s[p_0] &= \kappa (p^-_0 - p^+_0)- B(\mathcal{L}_s[p_0])^2, \\
   \mathcal{L}_{s-1}[p_0] &= \kappa (p^-_0 - p^+_0) - B ( \mathcal{L}_{s-1}[p_0])^2,
\end{align}
\end{subequations}
where $\mathcal{L}_s = (s^3\text{d}_{xx}/3 + s^2 s_x \text{d}_x$.
At $x = 0$, the inlet and no-gradient condition is 
\begin{equation} \label{eq:bc_1}
    p_{0}^- = 1, \quad p_{0x}^- = 0,
\end{equation}
and at $x = 1$, the outlet and no-gradient condition is 
\begin{equation} \label{eq:bc_2}
    p_{0}^+ = 0, \quad p_{0x}^+ = 0.
\end{equation}
Second, we consider the V-shaped filter with separators.
We can solve \eqref{eq:mass_0} subject to \eqref{eq:noslip_0}--\eqref{eq:sheet_2_0} in $D_1$ to give 
\begin{subequations} \label{eq:v_per_3}
\begin{align}
    u_0 &= p_{0x}(y - s)y/2, \\
    v_0 &= - p_{0xx}y^3 / 6 + (p_{0xx}s/4 + p_{0x} s_x/4)y^2, 
\end{align}
\end{subequations}
and in $D_2$ to give 
\begin{subequations} \label{eq:v_per_4}
\begin{align}
    u_0 &= p_{0x}(y^2 - (1+s)y + s)/2, \\
    v_0 &= p_{0xx}(-y^3/6 + (1+s)y^2/4 - sy/2 - (1 - 3s)/12) + p_{0x}s_x (y^2/2-y + 1/2)/2.
\end{align}
\end{subequations}
Substituting the velocity field \eqref{eq:v_per_3}--\eqref{eq:v_per_4} into \eqref{eq:sheet_1_0} along $y = s(x)$ gives 
\begin{subequations} \label{eq:bvp_3}
\begin{align}
   \mathcal{M}_s[p_0] &= \kappa (p^-_0 - p^+_0) - B(\mathcal{M}_s[p_0])^2, \\
   \mathcal{M}_{s-1}[p_0] &= \kappa (p^-_0 - p^+_0) - B(\mathcal{M}_{s-1}[p_0])^2,
\end{align}
\end{subequations}
subject to \eqref{eq:bc_1}--\eqref{eq:bc_2}, where $\mathcal{M}_s = \mathcal{L}_s/4$.
As $u_0 = u_0(x, \, y)$ due to the long-wave approximation, the flow rate through the half period \eqref{eq:flux_0} in the configuration without separators simplifies to
\begin{equation} \label{eq:flux_1}
    Q \approx W \int_{y = 0}^{1} u_0 \, \text{d} y =  -Wp_{0x}/3,
\end{equation}
and in the configuration with separators simplifies to
\begin{equation} \label{eq:flux_2}
    Q \approx W \int_{y = 0}^{1} u_0 \, \text{d} y = -Wp_{0x}/12.
\end{equation}
\update{Equation~\eqref{eq:flux_1} corresponds to the symmetric, separator-free configuration, where a shear-free condition at each half-period permits a higher parabolic velocity profile with mean coefficient $1/3$. 
Equation~\eqref{eq:flux_2} applies when separators are introduced, imposing no-slip boundaries at each half period, which increase shear and reduce the mean velocity coefficient to $1/12$. 
This difference, illustrated schematically in Fig.~\ref{fig:not_seperated}(a--d), highlights how the additional shear in the separator case lowers the effective flow rate. 
}
Under the long-wave assumption, the system parameters are reduced to just $W$, $\kappa$, $B$ and $\eta$; the goal in the following section is to identify the values that maximise $Q$ and $q$. 

\subsection{Validation of numerical solution}

\begin{figure*}[t!]
    \centering
  \includegraphics[width=.5\linewidth]{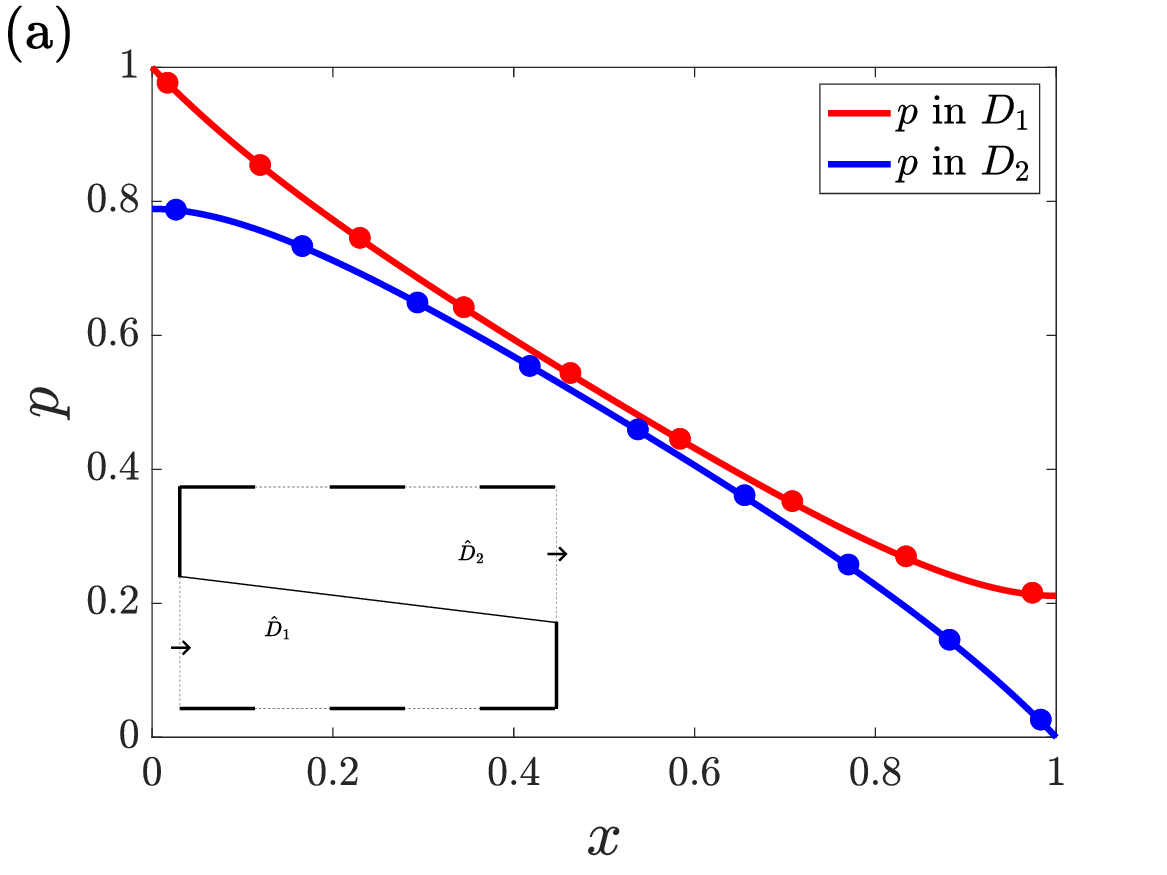}\includegraphics[width=.5\linewidth]{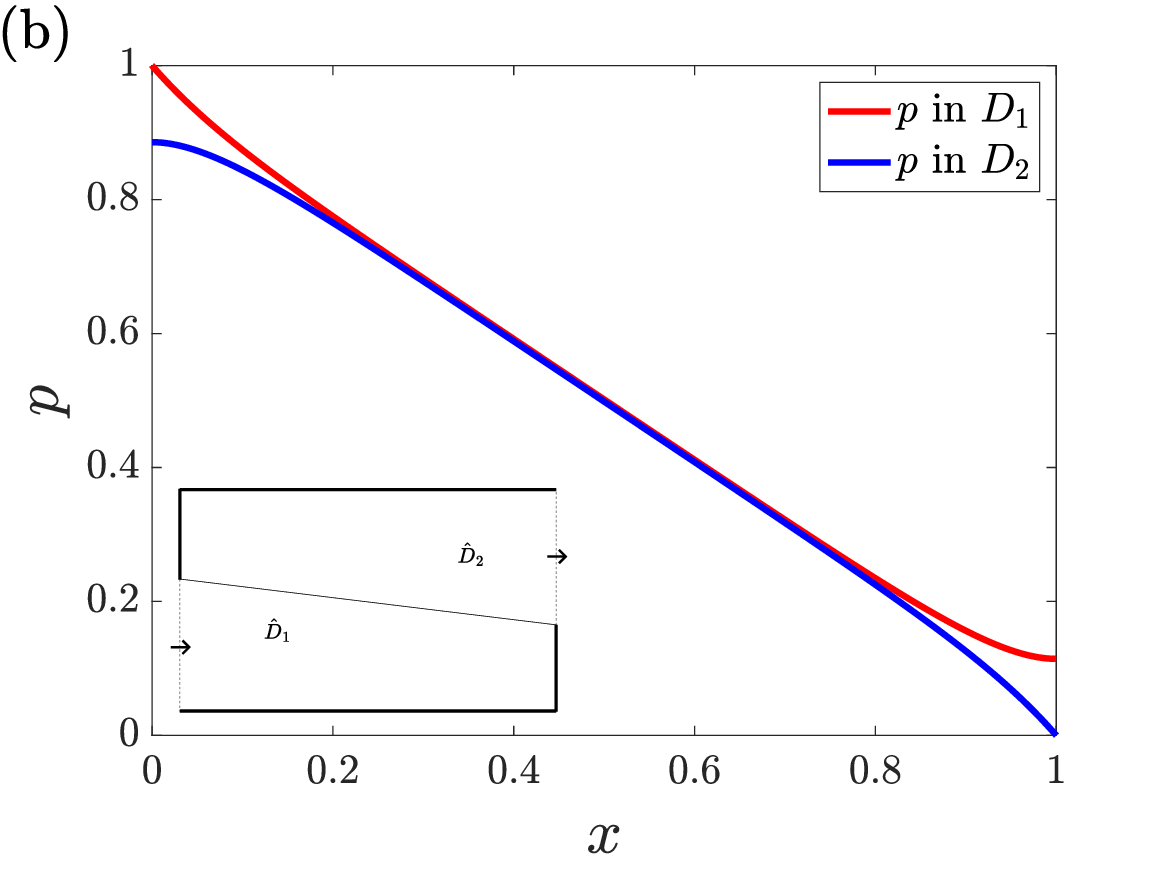}
    \caption{(a) Comparison of the pressure field in~\cite{pereira2021optimising} (symbols) with the long-wave model \eqref{eq:bvp_1} (lines) for a V-shaped filter without separators. 
    (b) The pressure field in the long-wave model \eqref{eq:bvp_3} for a V-shaped filter with separators.
    In (a--b), the location of the filter sheet is prescribed by $s(x) = a + \beta(1/2 - x)$, with $\beta = 0.05$ and $a = 0.5$ as illustrated in the inset, \update{such that the non-dimensional half-period length and height are unity and the permeance $\kappa = 1$. 
    }
    }
    \label{fig:Periera}
\end{figure*}

To validate the numerical solution to \eqref{eq:bvp_1}--\eqref{eq:bvp_3}, we compare the long-wave model's pressure field to that of~\cite{pereira2021optimising} for a filter without separators. 
Fig.~\ref{fig:Periera}(a) shows agreement, where the numerical solution is shown with solid lines and the results from~\cite{pereira2021optimising} are represented by symbols.  
In~\cite{pereira2021optimising}, the filter sheet did not extend across the entire half-period, and the model enforces no-slip and no-penetration regions above and below the inlet and outlet. 
Using the same parameters from~\cite{pereira2021optimising} ($\beta = 0.05$, $a = 0.5$ and $\kappa = 1$), the filter sheet is located at $s(x) = a + \beta(1/2 - x)$, which terminates at $s(0) = 0.525$ and $s(1) = 0.475$, \update{such that the non-dimensional half-period length and height are unity and the permeance $\kappa = 1$.} 
In Fig.~\ref{fig:Periera}(b), we present the pressure field obtained from the long-wave model for the same filter geometry without separators.
This allows for a direct comparison with the results from~\cite{pereira2021optimising} (Fig.~\ref{fig:Periera}a). 
While the pressure fields for filters with and without separators are similar, variation in the flow field and flow rate arises due to differences in boundary conditions at $y = 0,\,1$.
Following this validation step, we move on to explore filter sheets that span the full height of the pleat, with $s(0) = 1$ and $s(1) = 0$. 



\subsection{Permeability and removal efficiency}

Accurate permeability estimates are crucial for predicting the flow rate and optimising filter design.
The Kozeny--Carman model~\cite{kozeny1927ueber, carman1937fluid}, widely used for estimating permeability in porous media, is given by 
\begin{equation} \label{eq:kc} 
    \hat{k} = \frac{C \hat{d}^2 \phi^3}{(1-\phi)^2},  
\end{equation}  
where $C$ is a fitting coefficient determined from~\cite{del2002air, mrad2021local, zhang2022operating, al2011effect}.
Precise removal efficiency values are also critical for estimating filter effectiveness, particularly in environments requiring high air quality standards.
We model $\eta$ using an exponential model similar to the single-fibre efficiency formula described in~\cite{hinds2022aerosol},
\begin{equation}\label{eq:exp}
    \eta = r\left(1 - \exp\left(-\frac{E}{\hat{k}}\right)\right)
\end{equation}
where $E$ is a fitting coefficient determined from~\cite{del2002air, mrad2021local, zhang2022operating, al2011effect}.
The coefficient $r$ is equal to unity for particulate filtration and equal to the net probability of collision with an active site and conversion for catalytic removal.
\update{The constants $C$ and $E$ were determined by nonlinear least-squares regression using four experimental datasets~\cite{del2002air, mrad2021local, zhang2022operating, al2011effect}. 
For the permeability~\eqref{eq:kc}, $C$ was estimated by minimising 
\begin{equation}
    \text{SSE}_C = \sum_{i=1}^{4}\left(\hat{k}^{\mathrm{exp}}_i - \hat{k}^{\mathrm{model}}_i(C)\right)^2,
\end{equation}
and for the efficiency~\eqref{eq:exp}, $E$ was obtained from 
\begin{equation}
    \text{SSE}_E = \sum_{i=1}^{4}\left(\eta^{\mathrm{exp}}_i - \eta^{\mathrm{model}}_i(E)\right)^2.
\end{equation}
Optimisation was performed using MATLAB; the resulting values were $C = 0.07$ and $E = 2.1\times10^{-8}$~m$^2$. 
}

\section{Results} \label{sec:results}

\subsection{Experiments}

\begin{figure*}[t!]
    \begin{minipage}{0.49\linewidth}
    \includegraphics[width=.5\linewidth]{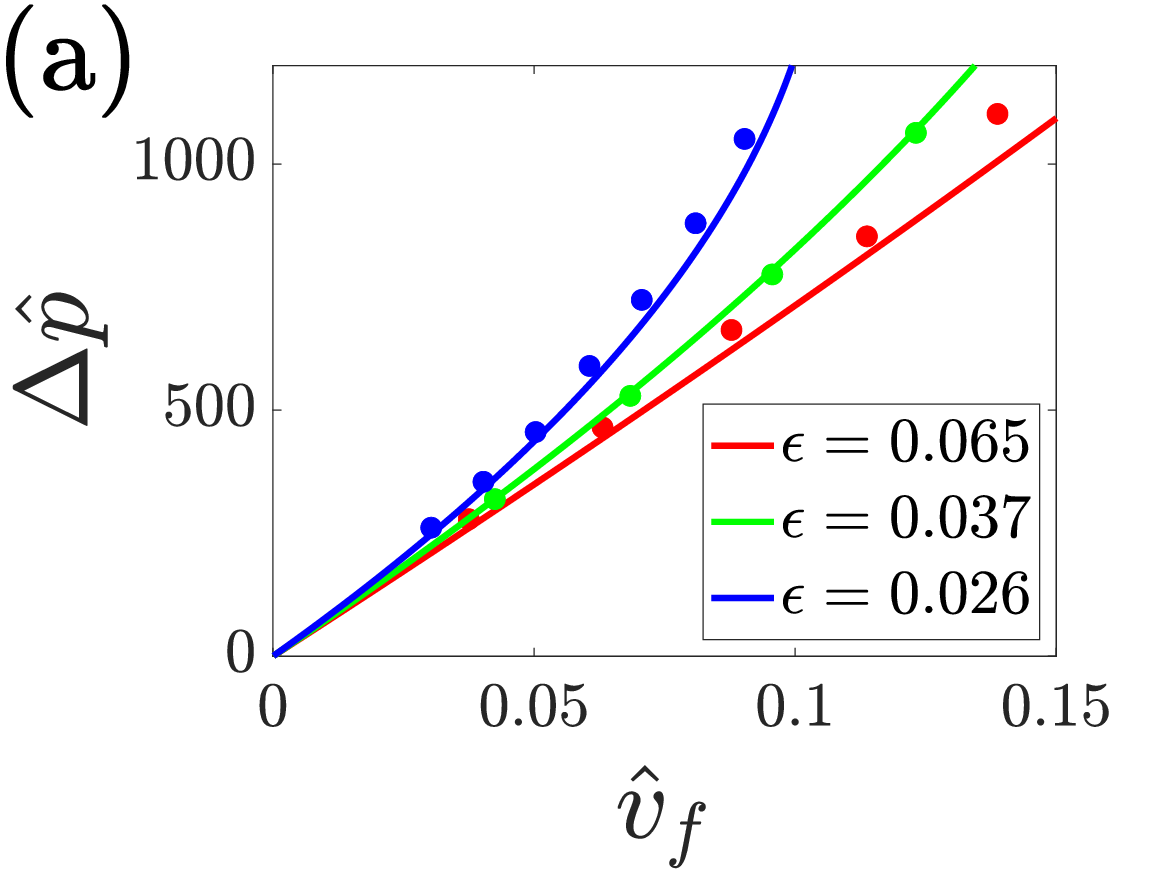}\includegraphics[width=.5\linewidth]{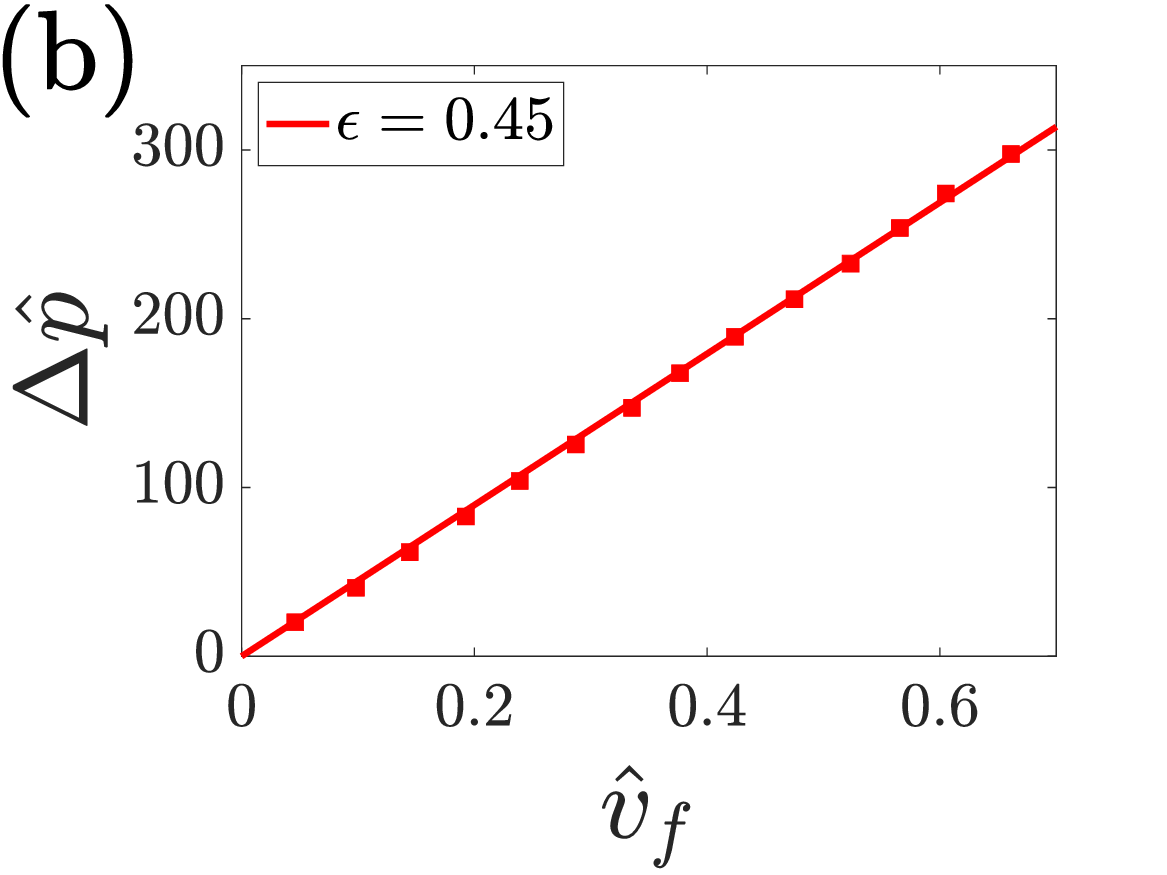}\\\includegraphics[width=.5\linewidth]{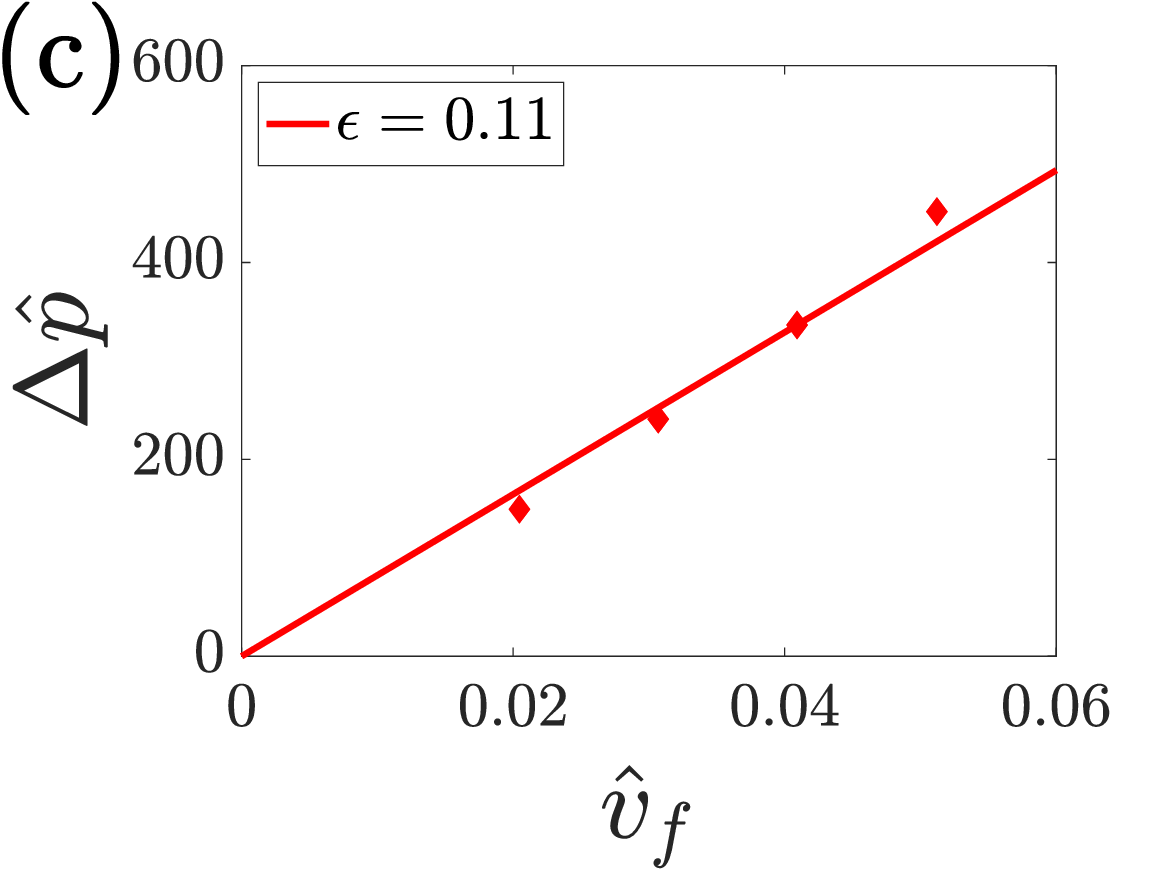}\includegraphics[width=.5\linewidth]{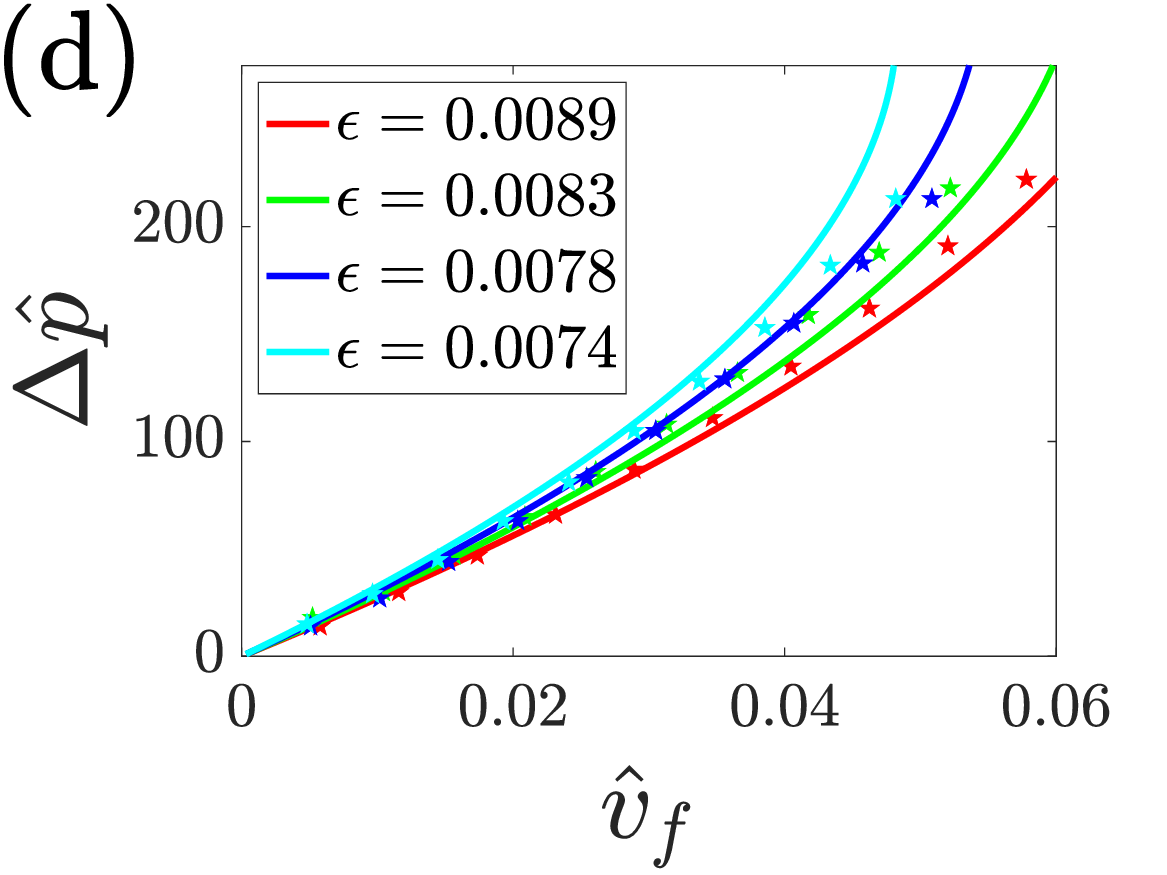}
    \end{minipage}
    \begin{minipage}{0.49\linewidth}
    \vspace{-.3cm}\includegraphics[width=\linewidth]{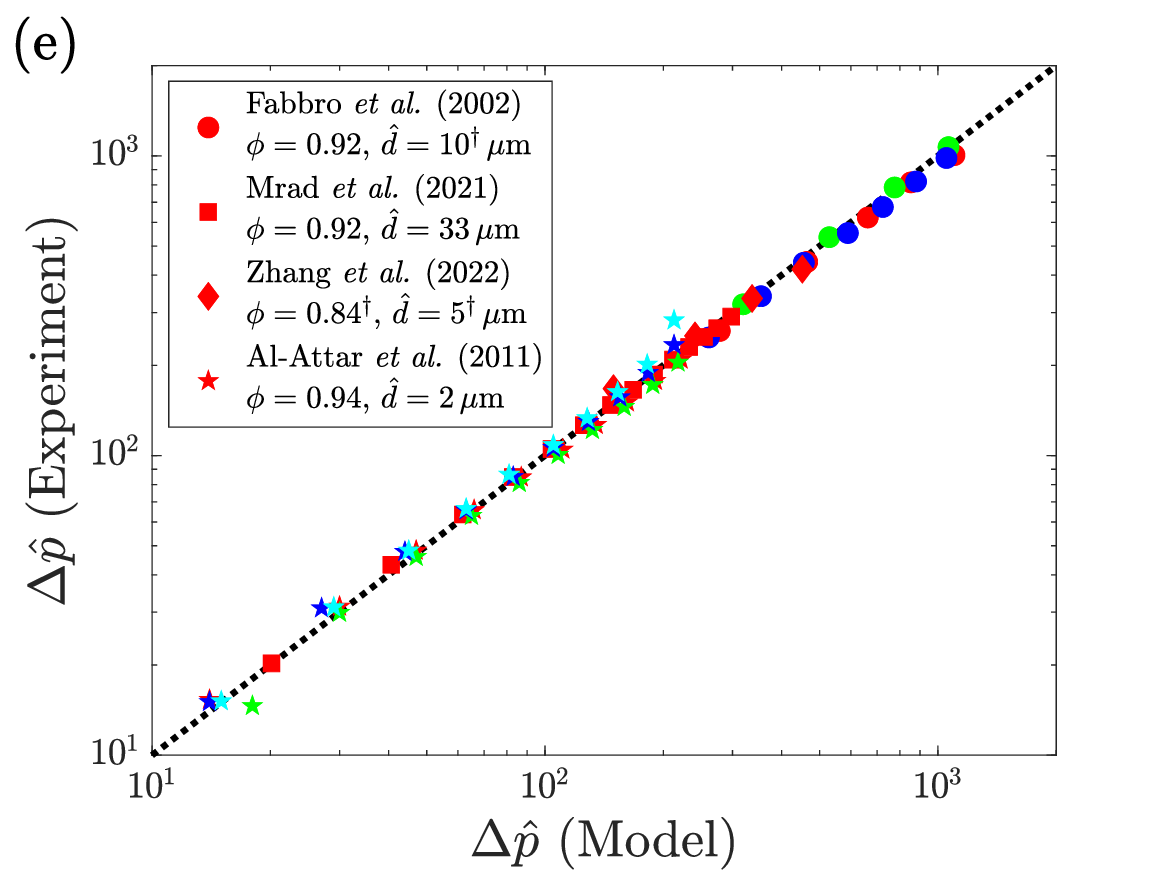}
    \end{minipage}
    \caption{Comparison of the pressure drop ($\Delta \hat{p}$) with the filtration velocity ($\hat{v}_f$) predicted using the long-wave model \eqref{eq:bvp_1} (solid lines) and experiments (symbols) in (a)~\cite{del2002air} for $\epsilon = 0.065$, 0.037, 0.026,
    (b)~\cite{mrad2021local} for $\epsilon = 0.45$, (c)~\cite{zhang2022operating} for $\epsilon = 0.11$ and (d)~\cite{al2011effect} for $\epsilon = 0.0089$, 0.0083, 0.0078, 0.0074. 
    (e) Comparison of $\Delta \hat{p}$ with $\hat{v}_f$ predicted using the long-wave model \eqref{eq:bvp_1} (solid lines) and~\cite{del2002air, mrad2021local,zhang2022operating,al2011effect} (symbols), \update{where porosity $\phi$ and fibre diameter $\hat{d}$ are listed for each dataset. 
    The superscript$^\dagger$ implies either that these values have been assumed or they have been calculated based on available data and \eqref{eq:kc}--\eqref{eq:exp}. 
    }}
    \label{fig:all}
\end{figure*}

To assess the accuracy of the long-wave model, we compare its predictions with experimentally observed relationships between the pressure drop ($\Delta \hat{p}$) and filtration velocity ($\hat{v}_f$) in V-shaped filters. 
\update{The validation parameters were selected from four laminar experimental datasets to ensure that the long-wave model accurately reproduces flow and pressure distributions across diverse filter configurations.} 
These experiments are detailed in Table \ref{tab:experiments}, where we exclude studies exhibiting turbulence or dust clogging for future work.
\update{The long-wave model behaviour varies with pressure drop and Reynolds number for different applications. 
As discussed in Section~\ref{sec:formulation}, for around $Re < 50$ Darcy behaviour dominates and for around $Re > 500$ inertial effects become significant~\cite{shin2022application}.
The Forchheimer correction (\ref{app:comp}) improves agreement for $Re = O(100)$ and the laminar model remains valid up to approximately $Re < 2000$~\cite{mcpherson2009subsurface} 
}
We compare the long-wave model's predictions with four experimental datasets in Fig.~\ref{fig:all}.
In Fig.~\ref{fig:all}(a), Fabbro \textit{et al.}~\cite{del2002air} measured the relationship between $\Delta \hat{p}$ with $\hat{v}_f$, showing agreement between their experimental results and the long-wave model with a permeability of $\hat{k} \approx 2\times10^{-9}$ m$^2$ \update{and an average inertial resistance of $\hat{\beta}_\text{mean} \approx 1.4\times10^{-4}$ s/m.} 
Since the study referenced nuclear applications, we assumed standard filter values typical for such settings: $\hat{d} = 10$ $\mu$m and $\eta = 99.97$\% for particles with a diameter of $\hat{d}_p \geq 0.3$ $\mu$m, consistent with similar configurations in Table \ref{tab:experiments}. 
At lower $\Delta \hat{p}$ values, the long-wave model closely predicts the experimental trends. 
Deviations emerge at higher pressure drops, likely due to strong inertial effects that are not accounted for in the model.

\begin{table}
    \centering
    \renewcommand{\arraystretch}{1.3} 
    \setlength{\tabcolsep}{4pt} 
    \scriptsize
    \begin{tabular}{lccccccc}
        \toprule
        Authors & Velocity ($\hat{v}_f$) & Geometry ($\epsilon$, $\hat{H}$) & Reynolds ($Re$) & Structure ($\phi$, $\hat{d}$) & Permeability ($\hat{k}$) & Efficiency ($\eta$) \\
        Fabbro \textit{et al.} \cite{del2002air} & 0.0746 & 0.0386 \quad 0.0016 & $4\times10^2$ & 0.92 \quad 10$^\dagger$ & $1 \times 10^{-9}$$^\dagger$ & $>$99$^\dagger$ \\
        Fabbro \textit{et al.} \cite{del2002air} & 0.0853 & 0.0359 \quad 0.0014 & $4\times10^2$ & 0.88 \quad 100$^\dagger$ & $4 \times 10^{-8}$$^\dagger$ & 50--55$^\dagger$ \\
        Mrad \textit{et al.} \cite{mrad2021local} & 0.3315 & 0.4464 \quad 0.0125 & $1\times10^{3}$ & 0.92 \quad 33 & $2 \times 10^{-8}$$^\dagger$ & 50--55 \\
        Zhang \textit{et al.} \cite{zhang2022operating} & 0.0358 & 0.1092 \quad 0.0155 & $6\times10^2$ & 0.84$^\dagger$ \quad 5$^\dagger$ & $1 \times 10^{-9}$$^\dagger$ & 99.97 \\
        Theron \textit{et al.}$^*$ \cite{theron2017numerical} & 0.449 & 0.3195 \quad 0.0064 & $2\times10^{3}$ & 0.92 \quad 13.8 & $2 \times10^{-9}$$^\dagger$ & $>$99$^\dagger$ \\
        Tronville \textit{et al.}$^*$ \cite{tronville2003minimization} & 1.25 & 0.1280 \quad 0.0032 & $4\times10^{3}$ & 0.92$^\dagger$ \quad 1$^\dagger$ & $1 \times 10^{-10}$$^\dagger$ & $>$99$^\dagger$  \\
        Youssef \textit{et al.} \cite{youssef2016experimental} & 0.2 & 0.125 \quad 0.0025 & $3\times10^2$ & 0.92$^\dagger$ \quad 1$^\dagger$ & $1\times10^{-10}$$^\dagger$ & $>$99$^\dagger$ \\
        Rebai \textit{et al.}$^*$  \cite{rebai2010semi} & 0.65 & 0.078 \quad 0.004 & $2\times10^3$ & 0.92$^\dagger$ \quad 15 & $4\times10^{-10}$$^\dagger$ & $>$99$^\dagger$ \\
        Al-Attar \textit{et al.} \cite{al2011effect} & 0.0287 & 0.008 \quad 0.003 & $3\times10^2$ & 0.94 \quad 2.1 & $6\times10^{-9}$$^\dagger$ & 94 \\
        \bottomrule
    \end{tabular}  
    \caption{Summary of experiments~\cite{del2002air, mrad2021local, zhang2022operating, theron2017numerical, tronville2003minimization, youssef2016experimental, rebai2010semi, al2011effect} on V-shaped filters without dust clogging: their average filter velocity ($\hat{v}_f$, m/s), aspect ratio ($\epsilon$, --), half-height ($\hat{H}$, m), Reynolds number ($Re$, --), porosity ($\phi$, --), fibre diameter ($\hat{d}$, $\mu$m), permeability ($\hat{k}$, m$^2$) and efficiency ($\eta$, --). 
    The superscript$^\dagger$ implies either that these values have been assumed or they have been calculated based on available data and \eqref{eq:kc}--\eqref{eq:exp}. 
    \update{The superscript$^*$ indicates studies for which ${Re} > 2000$, where the flow is considered turbulent~\cite{mcpherson2009subsurface}.}}
    \label{tab:experiments}
\end{table}

Moving on to a different experimental setup, Mrad \textit{et al.}~\cite{mrad2021local} reported $\Delta \hat{p}$ as a function of $\hat{v}_p$, with the long-wave model showing agreement in Fig.~\ref{fig:all}(b) when $\hat{k} \approx 2\times 10^{-8}$ m$^2$, where $\eta = 50$--$55$\% for $\hat{d}_p =1$ $\mu$m, $\phi = 0.92$ and $\hat{d} = 33$--$183$ $\mu$m. 
Next, Zhang \textit{et al.}~\cite{zhang2022operating} measured the pressure drop as a function of airflow rate $\hat{Q}$ (m$^3$/h), which we converted $\hat{Q}$ to $\hat{v}_f$ using a provided table to facilitate comparison. 
Fig.~\ref{fig:all}(c) shows agreement, with the long-wave model fitting their experimental data when $\hat{k} \approx 1\times 10^{-9}$ m$^2$ and $\phi = 0.84$, where $\eta = 99.97$\% for $\hat{d}_p =0.3$ $\mu$m and $\hat{d} = 5$ $\mu$m. 
Finally, Al-Attar \textit{et al.}~\cite{al2011effect} explored $\Delta\hat{p}$ as a function of $\hat{v}_p$ and $\epsilon$. 
Fig.~\ref{fig:all}(d) demonstrates the long-wave model’s correspondence with their experimental data with $\hat{k} \approx 6\times 10^{-9}$ m$^2$ \update{and $\hat{\beta}_\text{mean} \approx 3.6\times10^{-4}$ s/m}, where $\eta = 94$\% for $\hat{d}_p \geq 0.1$ $\mu$m, $\phi = 0.94$ and $\hat{d} = 2.1$ $\mu$m. 
Across all V-shaped filter datasets, the long-wave model successfully captures the effects of geometry, permeability and flow resistance, as summarised in Fig.~\ref{fig:all}(e).

\begin{figure*}[t!]
    \centering
    \includegraphics[width=.5\linewidth]{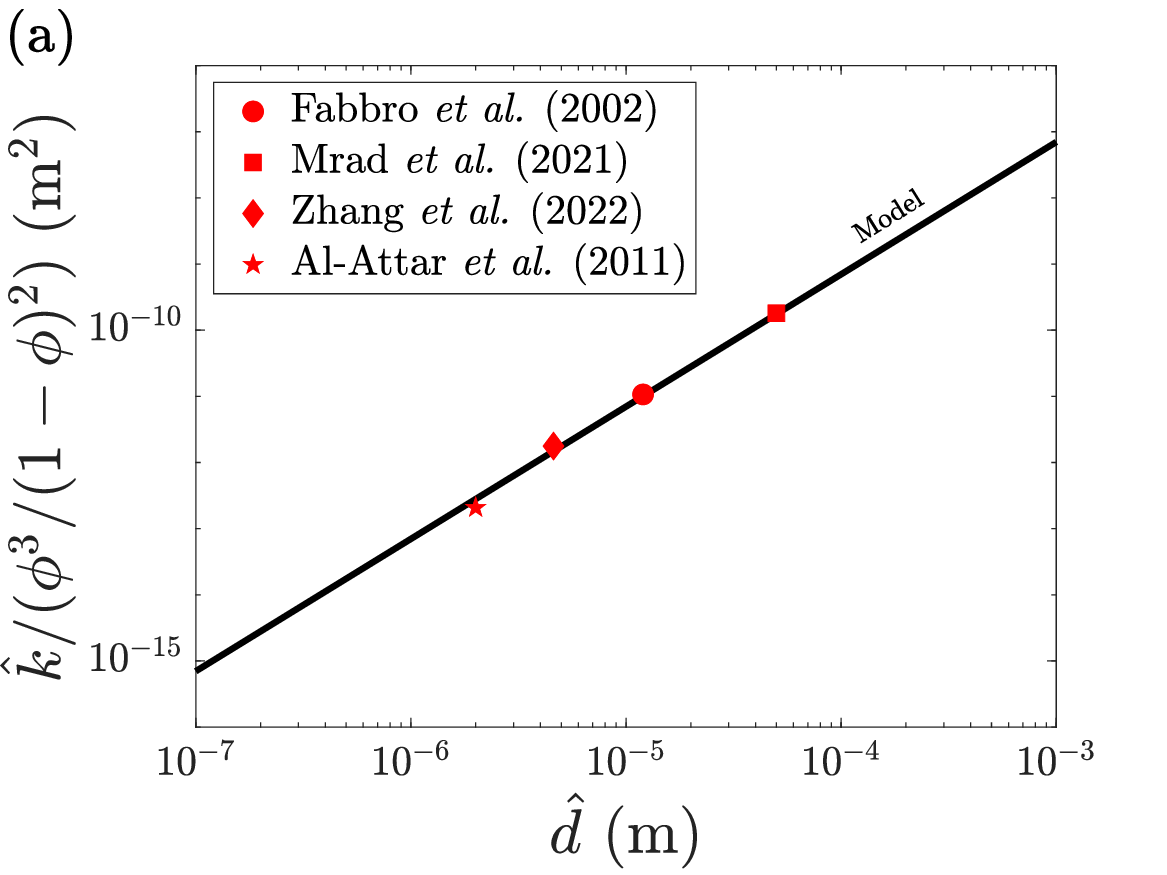}\includegraphics[width=.5\linewidth]{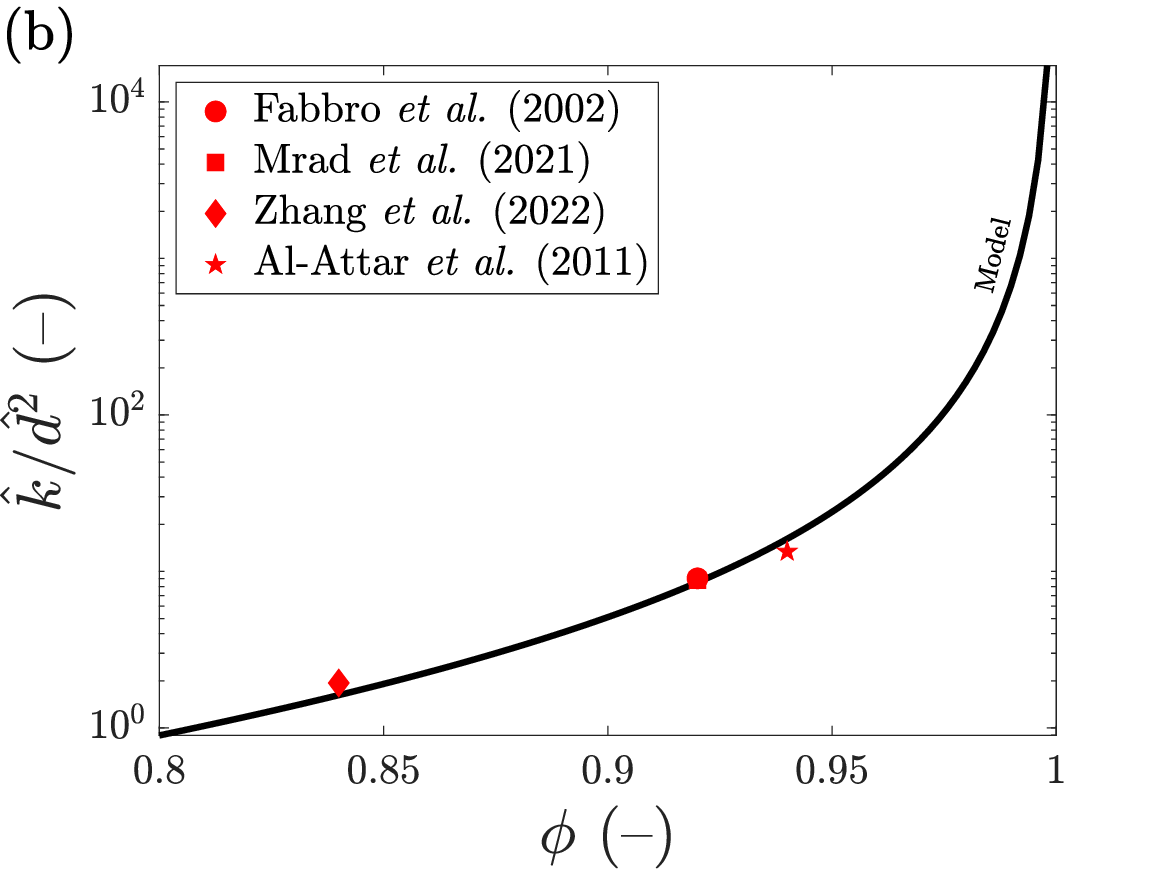}
    \caption{Comparison of the normalised permeability (a) $\hat{k} / (\phi^3/(1-\phi^2))$ and (b) $\hat{k} / \hat{d}^2$ with the average fibre diameter ($\hat{d}$) and porosity ($\phi$) using the Kozeny--Carman model \eqref{eq:kc} (solid lines) and experiments~\cite{del2002air, mrad2021local, zhang2022operating, al2011effect} (symbols).}
    \label{fig:kc}
\end{figure*}

Having demonstrated agreement between the long-wave model and the experimental data (Fig.~\ref{fig:all}), we next show that the permeability can be predicted from measurable filter-sheet properties, eliminating the need for case-by-case fitting.
Fig.~\ref{fig:kc} shows that the Kozeny--Carman model~\cite{kozeny1927ueber, carman1937fluid} offers a robust framework for estimating permeability, effectively capturing how variations in $\hat{d}$ and $\phi$ influence $\hat{k}$ across the experiments.
The Kozeny--Carman model predicts that permeability increases with both $\hat{d}$ and $\phi$, consistent with trends observed in the experimental data.
By collapsing the experimental trends related to fibre diameter and porosity into a single relation, the long-wave model demonstrates its broad applicability to V-shaped filter configurations, offering a unified approach to predicting filter performance. 

\begin{figure}[t!]
    \centering
    \includegraphics[width=.5\linewidth]{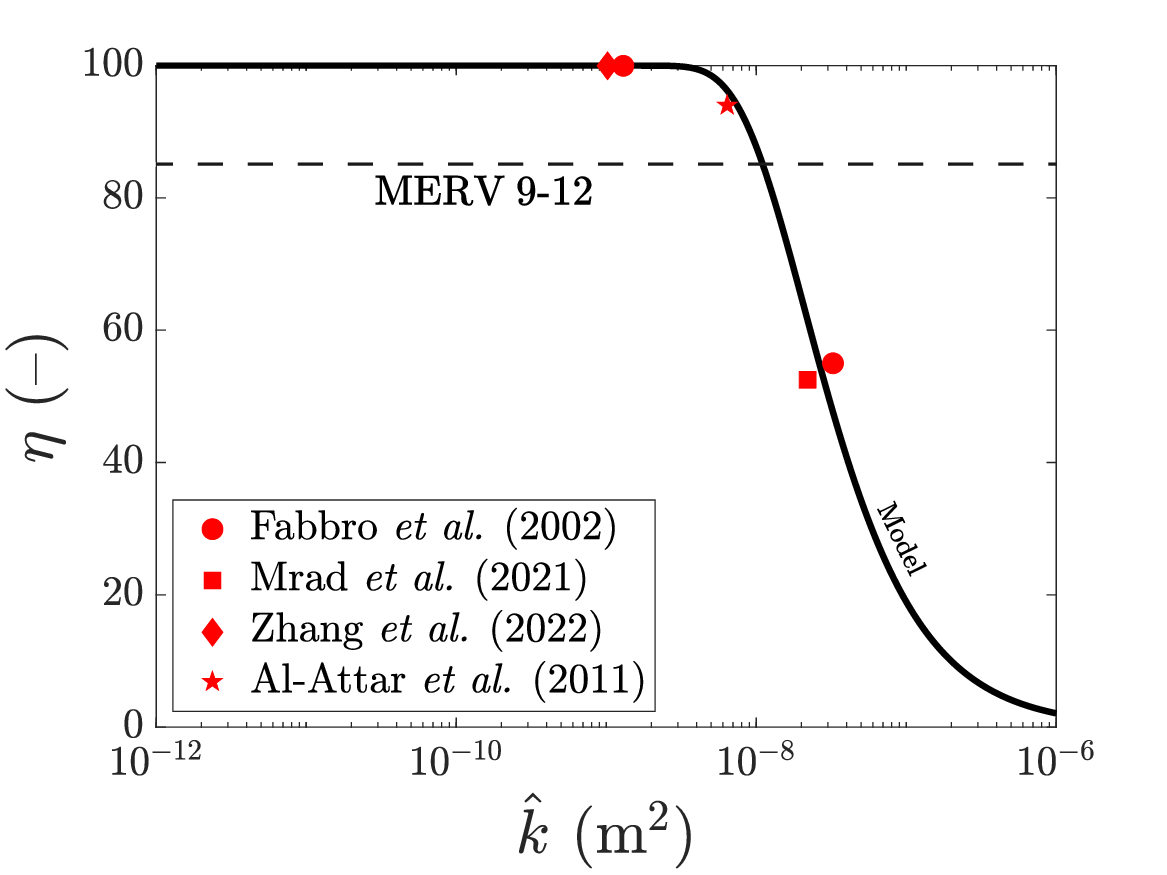}
    \caption{Comparison of the removal efficiency $\eta$ with the permeability $\hat{k}$ using the exponential model~\eqref{eq:exp} (solid line) and experiments~\cite{del2002air, mrad2021local, zhang2022operating, teng2022research} (symbols).
    Note that as $\hat{d}$ varies from $1\times10^{-6}$ in Fabbro \textit{et al.}~\cite{del2002air} to $6\times10^{-6}$ m in Fig.~\ref{fig:flux}(a), $Q$ increases but $\eta$ decreases from 99.97 to 29\%.}
    \label{fig:removal_efficiency}
\end{figure}

Since permeability plays a central role in filtration performance, we extend this framework to investigate how changes in the filter-sheet properties affect the removal efficiency ($\eta$) of the V-shaped filter for particles with diameter $\hat{d}_p \geq 0.3$ $\mu$m in Fig.~\ref{fig:removal_efficiency}.
The exponential model reflects the experimental trend that higher permeability leads to lower removal efficiency, as increased pore space reduces particle capture efficiency.
Thus, we have established a predictive framework that links measurable V-shaped filter properties to the permeability through the Kozeny–Carman model~\eqref{eq:kc}, the removal efficiency through the exponential model~\eqref{eq:exp} and flow rates through the long-wave model~\eqref{eq:bvp_1}--\eqref{eq:flux_2}.
When investigating changes in the flow rate due to catalytic coatings in the following section, we will ensure that the V-shaped filters meet a MERV rating of at least 9–12 for particles with $\hat{d}_p \geq 1$ $\mu$m, corresponding to a minimum efficiency reporting value of 85\%~\cite{ASHRAE2024}, as depicted by the black dashed line in Fig.~\ref{fig:removal_efficiency}.

\subsection{Flow rate} \label{subsec:flowrate}

\begin{figure*}[t!]
    \centering
  \includegraphics[width=.5\linewidth]{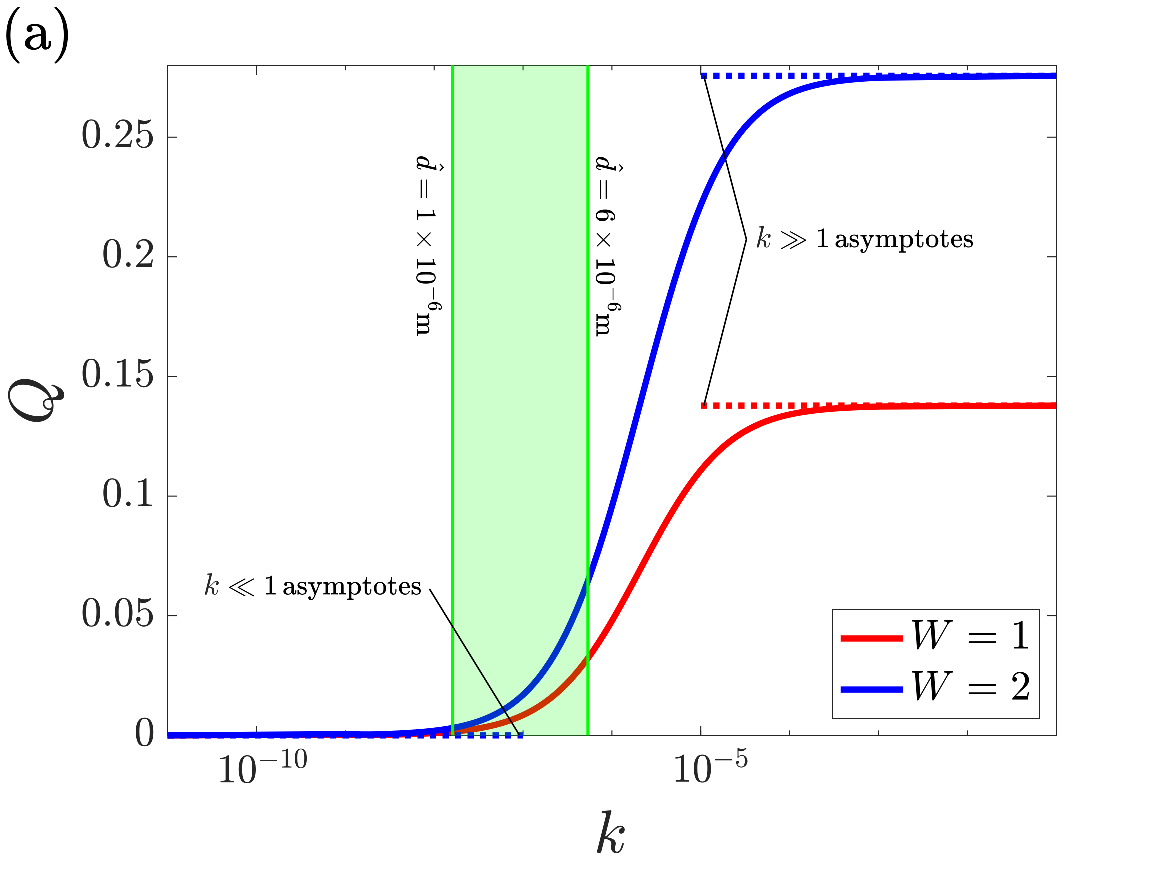}\includegraphics[width=.5\linewidth]{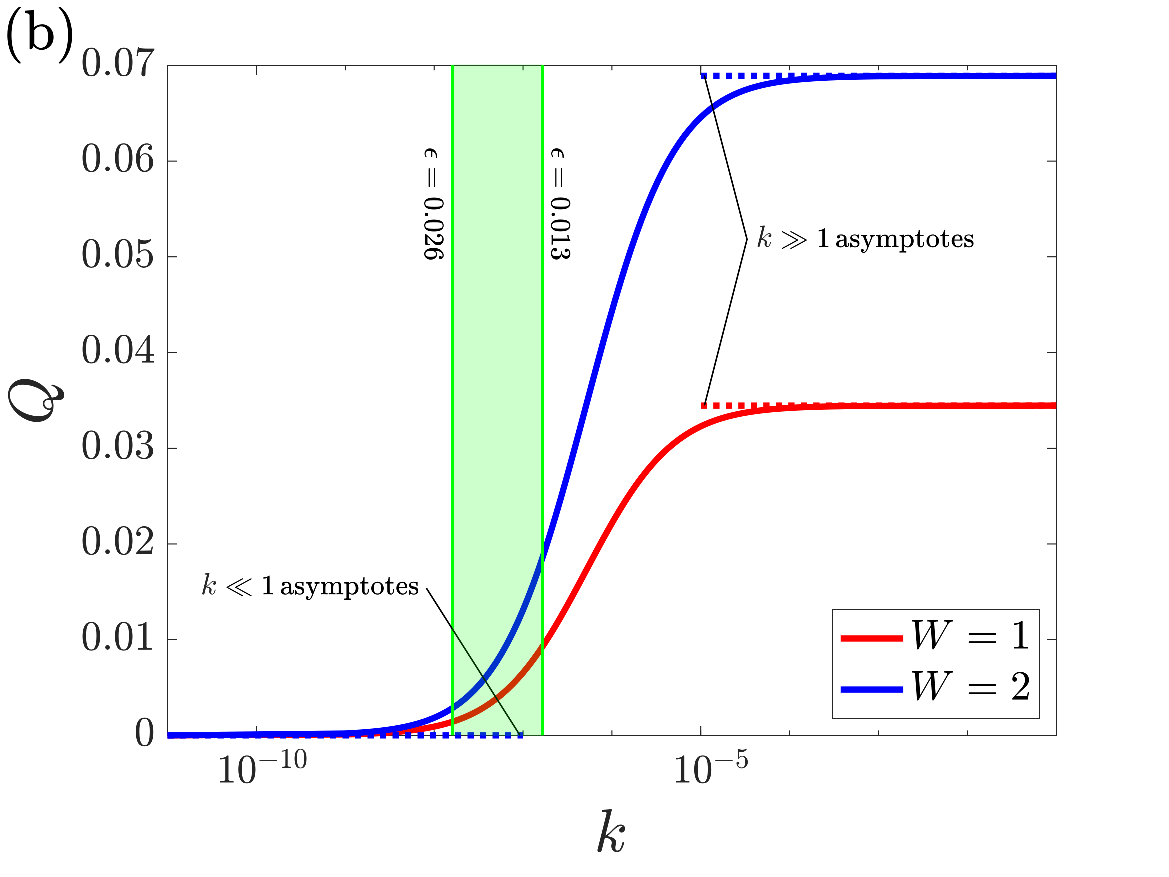}    \caption{The flow rate ($Q$) through the V-shaped filter (a) without separators and (b) with separators, for varying permeability ($k$) and filter widths ($W$), using (\ref{eq:bvp_1},\,\ref{eq:bvp_3}). 
  Dashed lines represent asymptotes derived in \ref{app:asym} for $k \gg 1$ and $k \ll 1$, and vertical lines represent the non-dimensional permeability measured in experiments in~\cite{del2002air} (green) for varying fibre diameter ($\hat{d}$) and aspect ratio ($\epsilon$).}
    \label{fig:flux}
\end{figure*}

Fig.~\ref{fig:flux} illustrates how the non-dimensional permeability ($k$) and filter width ($W$) influence the flow rate ($Q$) in V-shaped filters, providing insights into optimising filter performance. 
Separators reduce the maximum flow rate by a factor of four compared to configurations without separators.
\update{A parametric sweep over $\hat{H}$ $\hat{L}$, $\hat{\kappa}$ and $\hat{\mu}$ shows that the ratio $Q_{\mathrm{separators}}/Q_{\mathrm{no\text{-}separators}}$ varies between 0.29 and 0.25 for $\kappa > 1$, quantifying the accuracy of the factor-of-four reduction in the strong-permeability, long-wave limit. 
} 
\update{In this limit, where the filter resistance is negligible, the factor-of-four reduction arises from geometric confinement and remains consistent across geometries when $\kappa \gg 1$ and $\epsilon \ll 1$, as shown in~\ref{app:asym}. 
}
To ground the findings within experimental data, we highlight a permeability estimate from Table \ref{tab:experiments} (green), where $k = 2\times10^{-8}$, $\hat{d} = 1\times10^{-6}$ m and $\epsilon = 0.013$.

Fig.~\ref{fig:flux}(a) illustrates that increasing $\hat{d}$, e.g., by applying a catalyst coating with an average thickness of $5 \mu$m to the fibres that make up the filter sheet (assuming the catalyst is applied before the fibres are assembled; if applied after the filter is assembled, the relationship between $k$ and the filter’s structure will differ), leads to a higher $k$, which in turn increases $Q$ from $0.003$ to $0.07$ for $W=2$.
Larger fibres create wider pores, reducing the flow resistance and allowing for more efficient airflow.
However, increasing the permeability negatively impacts removal efficiency.
As $k$ increases, $\eta$ decreases from 99.97 to 29\%, although the total filtered airflow $\eta Q$ still increases from $0.003$ to $0.02$.
To maintain a removal efficiency of at least $85$\% (MERV 9–12) for the V-shaped filter from Table \ref{tab:experiments}, permeability must not exceed $\hat{k}<1\times10^{-8}$ m$^2$, limiting the increase in $\hat{d}$ to 2.5 times, compared to the sixfold increase shown in Fig.~\ref{fig:flux}(a).

Fig.~\ref{fig:flux}(b), examines how reducing $\epsilon$, e.g., by doubling the V-shaped filter length relative to the period half-height, affects permeability and flow rate.
This change increases $k$ and raises $Q$ from $0.003$ to $0.02$ for $W = 2$. 
Extending the fibre sheet increases its surface area, allowing for greater airflow and higher flow rates through the filter.
In this case, the increase in $k$ due to $\epsilon$ does not significantly impact $\eta$, as indicated by \eqref{eq:exp}.
Our findings highlight the intricate balance between permeability, flow rate and removal efficiency in V-shaped filters. 
While increasing fibre diameter or filter length can improve airflow, careful parameter selection is necessary to maintain high filtration performance.
These insights generalise to other parameters, including $\phi$, $\hat{t}$ and $\hat{L}$, each of which influences permeability and flow dynamics.

\subsection{Flow and pressure field} 

\begin{figure*}[t!]
    \centering
  \small (a) \hfill (b) \hfill (c) \hfill (d) \hfill \hfill \hfill \\ \includegraphics[width=.24\linewidth]{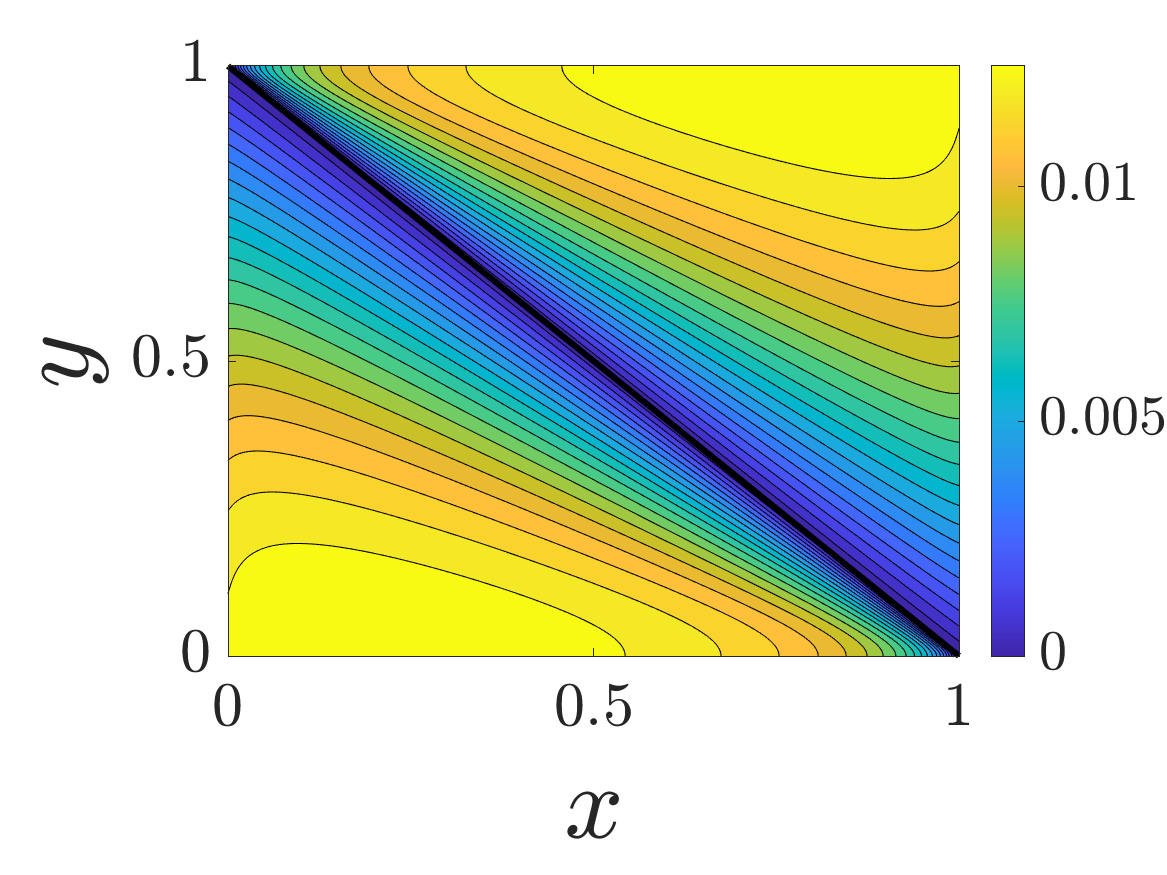}
  \includegraphics[width=.24\linewidth]{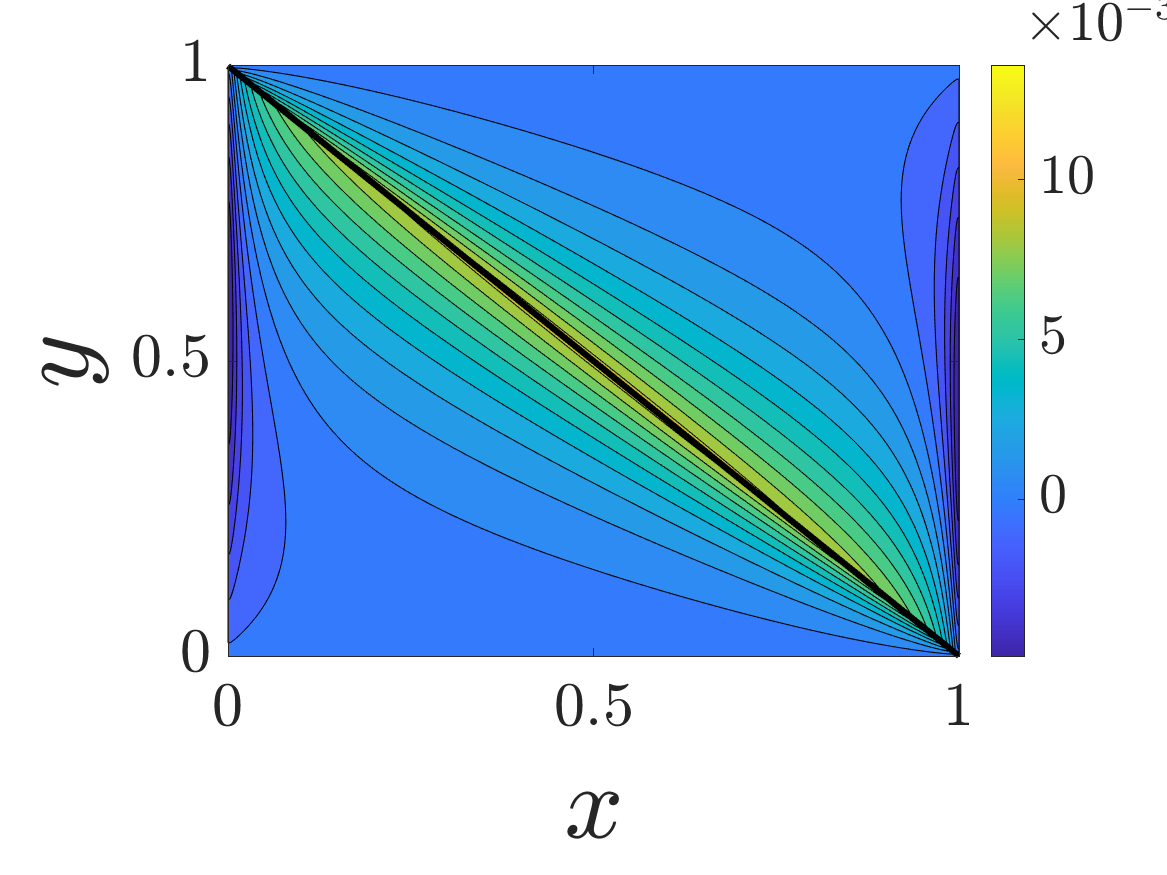}
  \includegraphics[width=.24\linewidth]{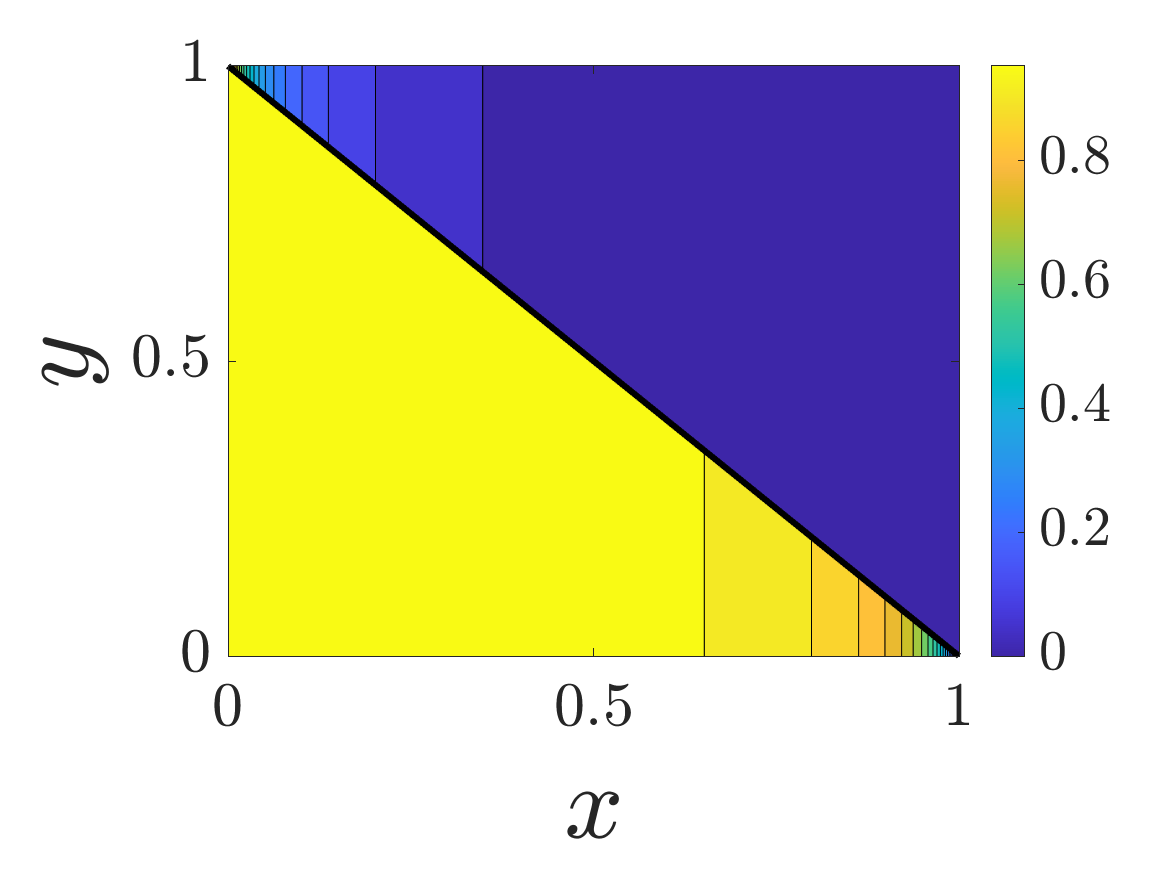}
  \includegraphics[width=.24\linewidth]{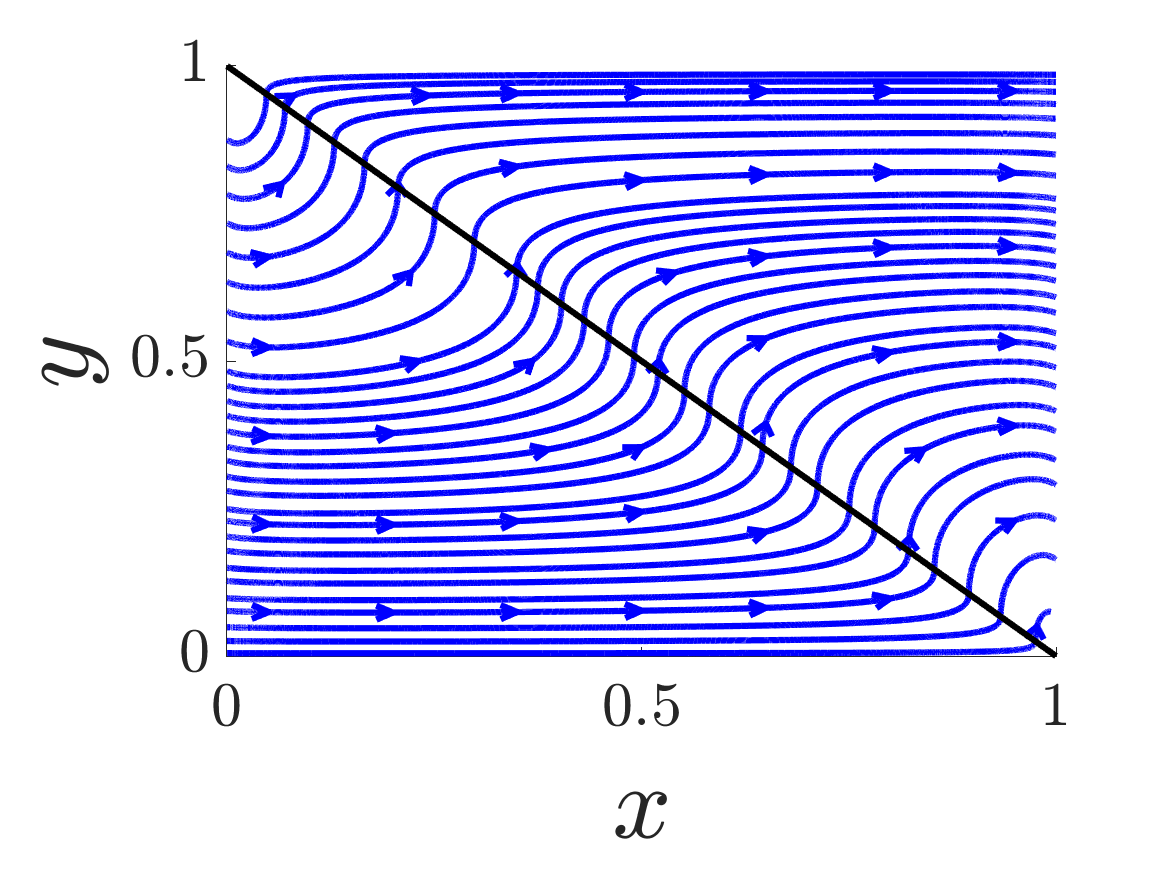}\\
  \vspace{-.5cm}(e) \hfill (f) \hfill (g) \hfill (h) \hfill \hfill \hfill \\ 
  \includegraphics[width=.24\linewidth]{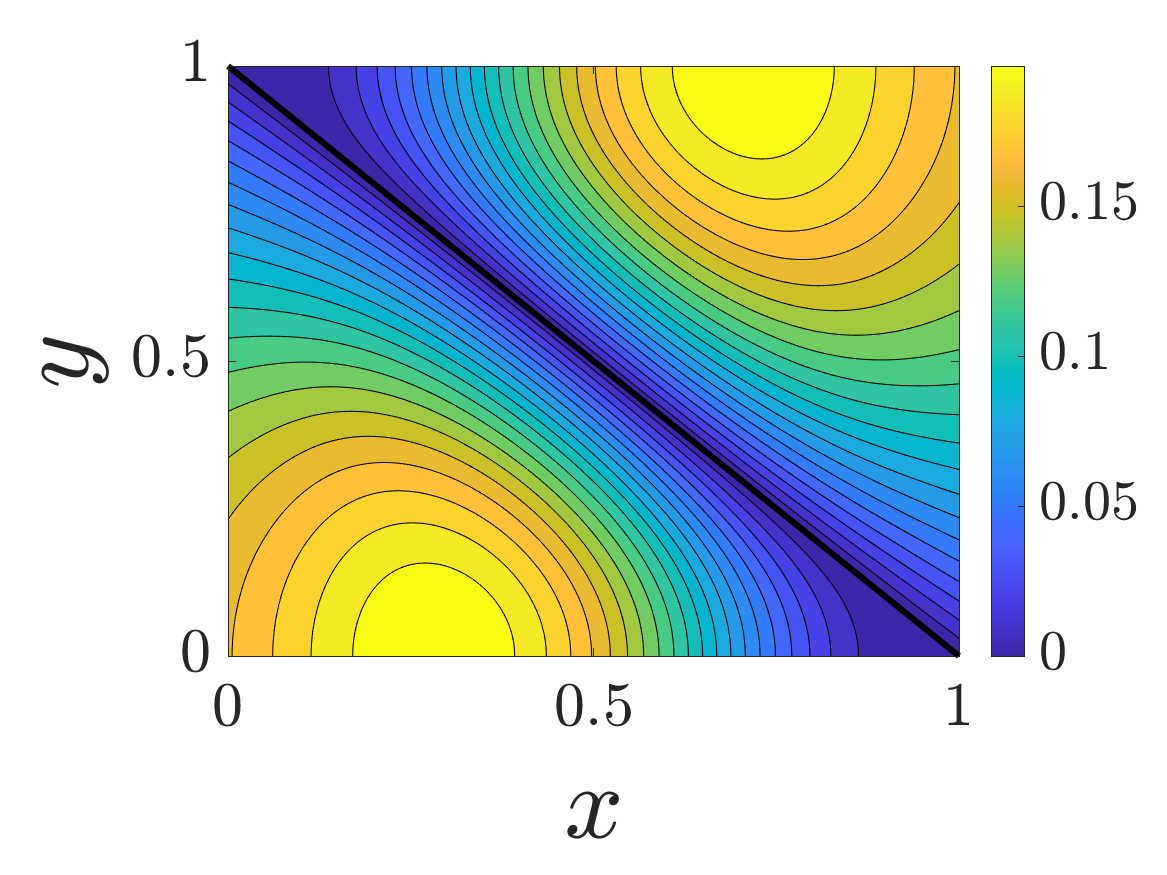}
  \includegraphics[width=.24\linewidth]{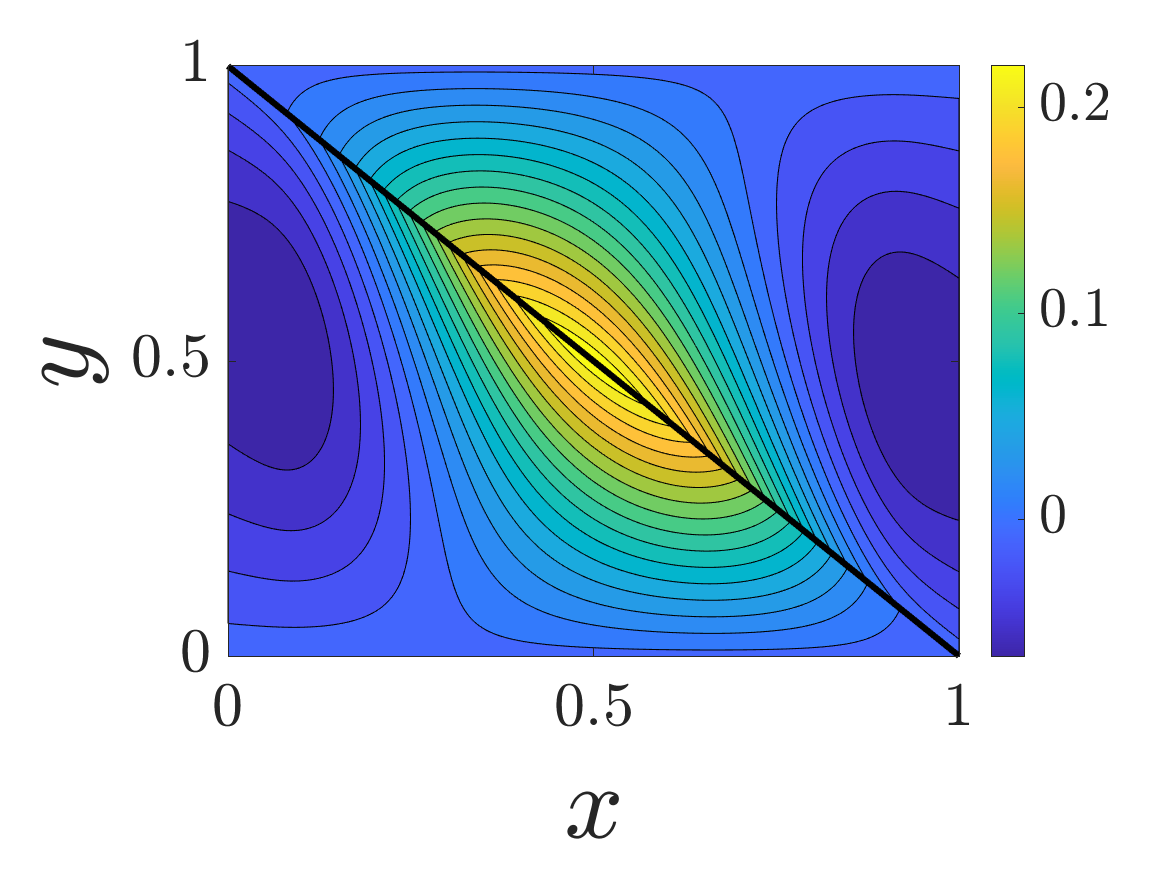}
  \includegraphics[width=.24\linewidth]{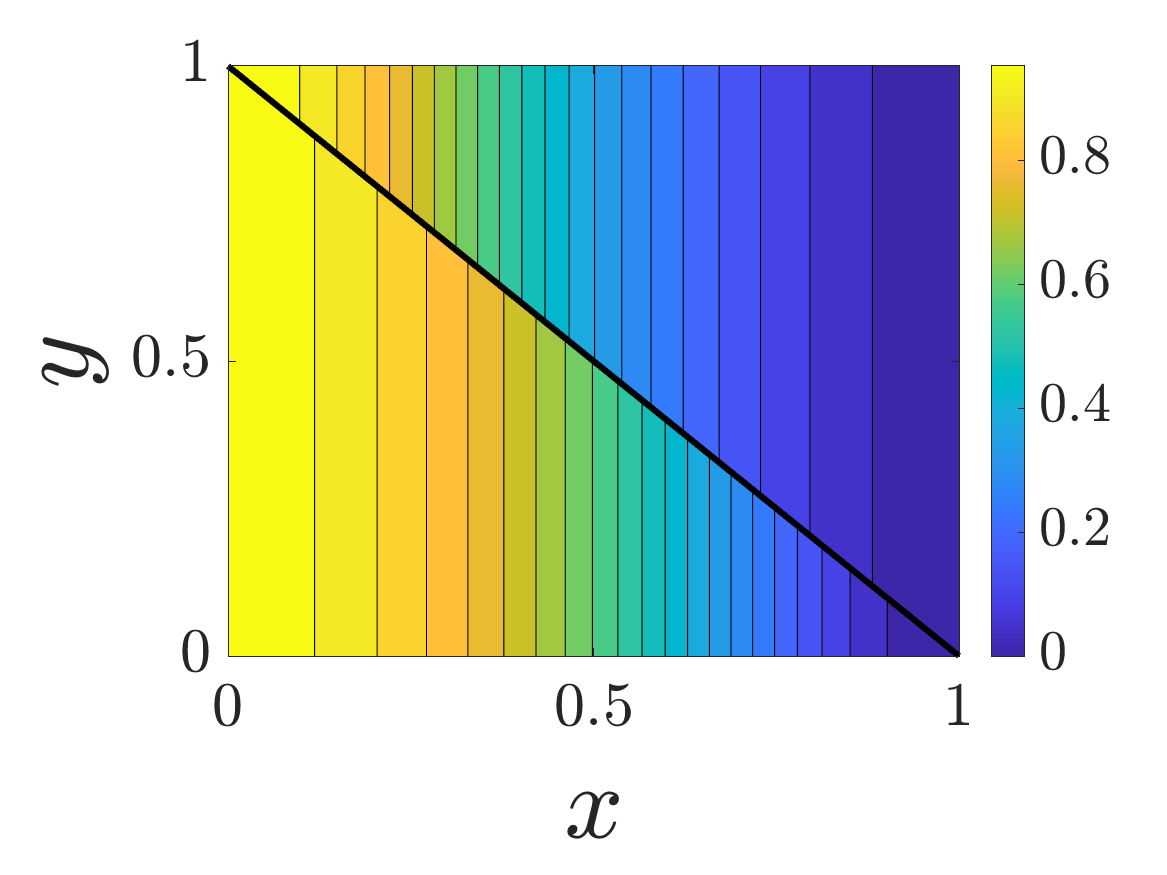}
  \includegraphics[width=.24\linewidth]{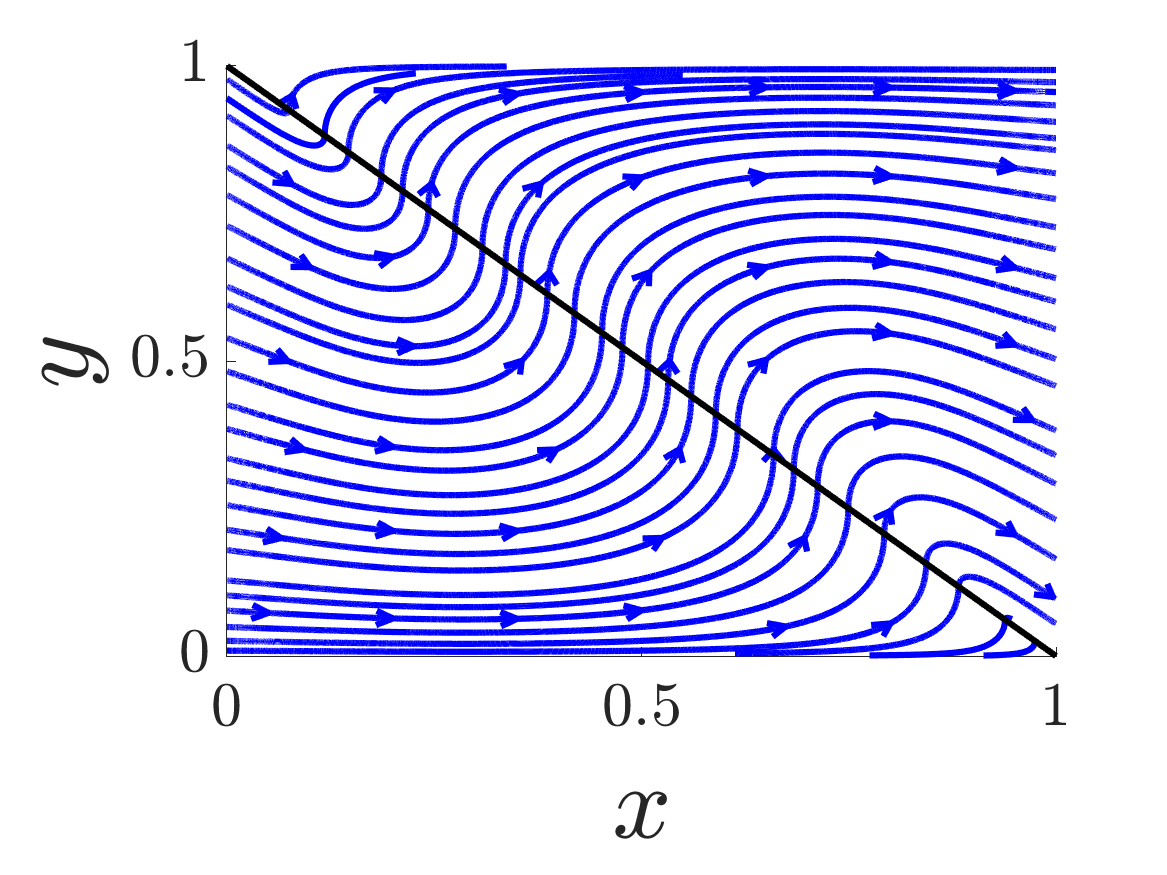}\\
    \vspace{-.5cm}(i) \hfill (j) \hfill (k) \hfill (l) \hfill \hfill \hfill \\ 
  \includegraphics[width=.24\linewidth]{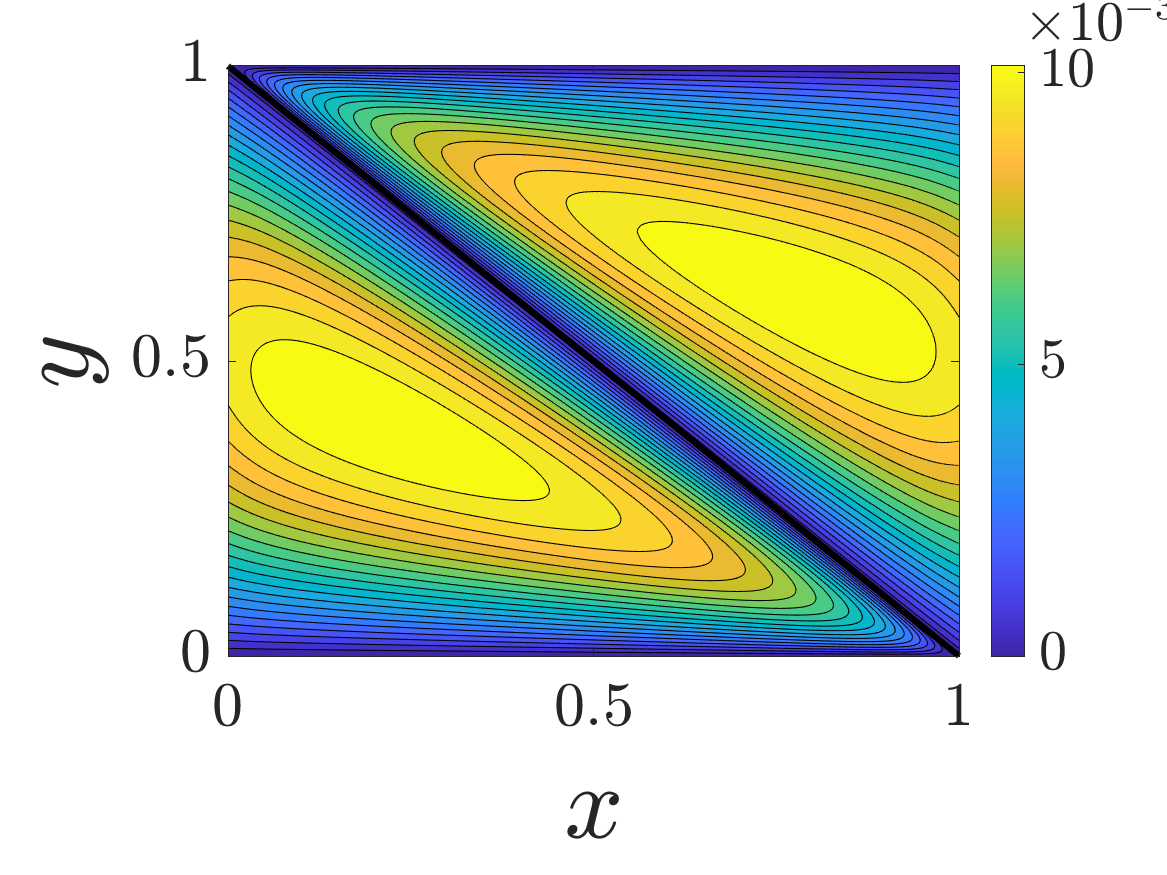}
  \includegraphics[width=.24\linewidth]{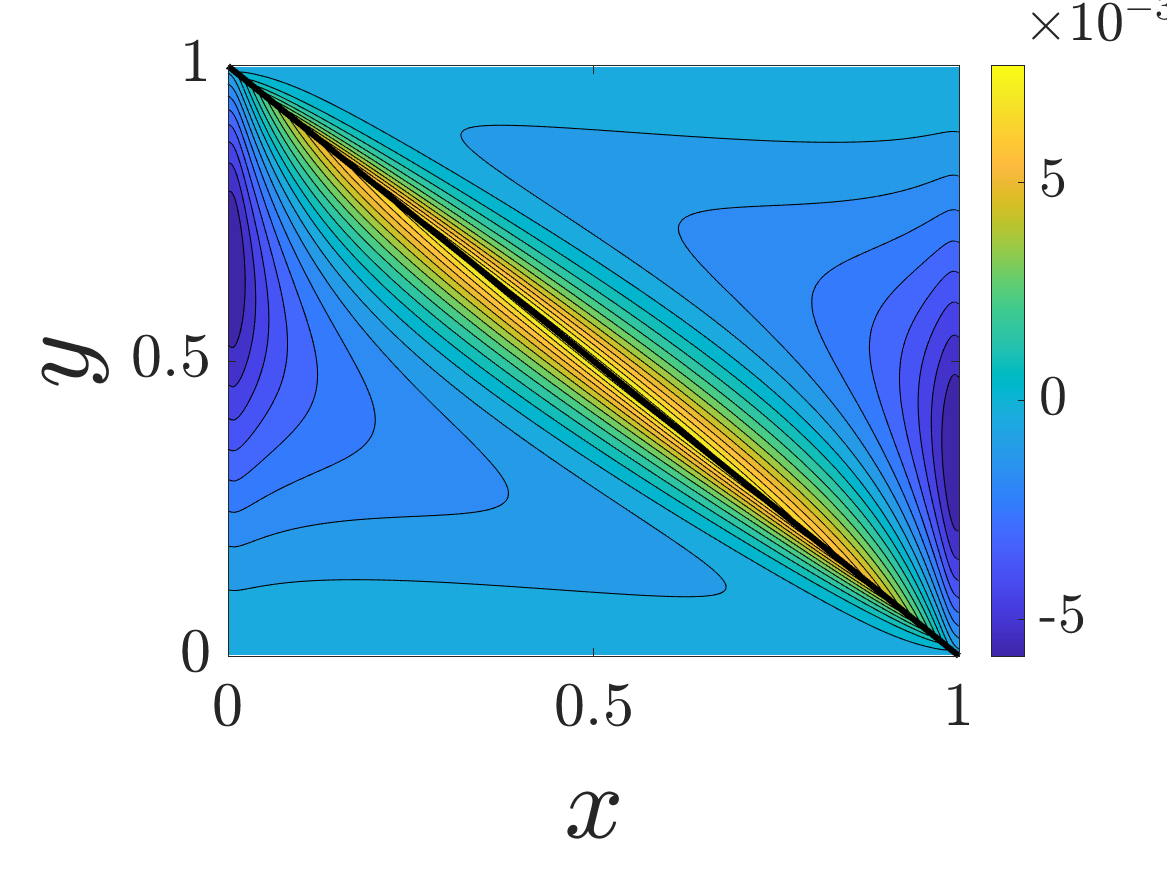}
  \includegraphics[width=.24\linewidth]{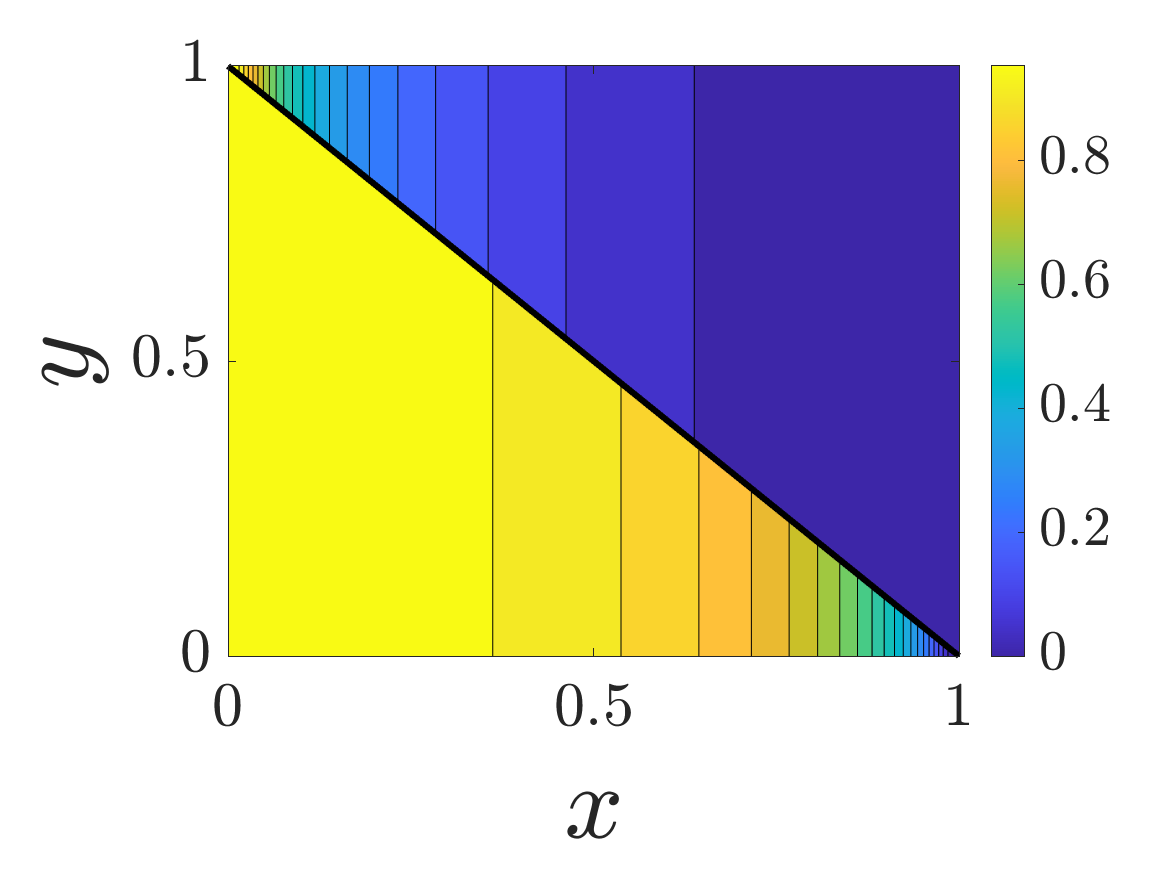}
  \includegraphics[width=.24\linewidth]{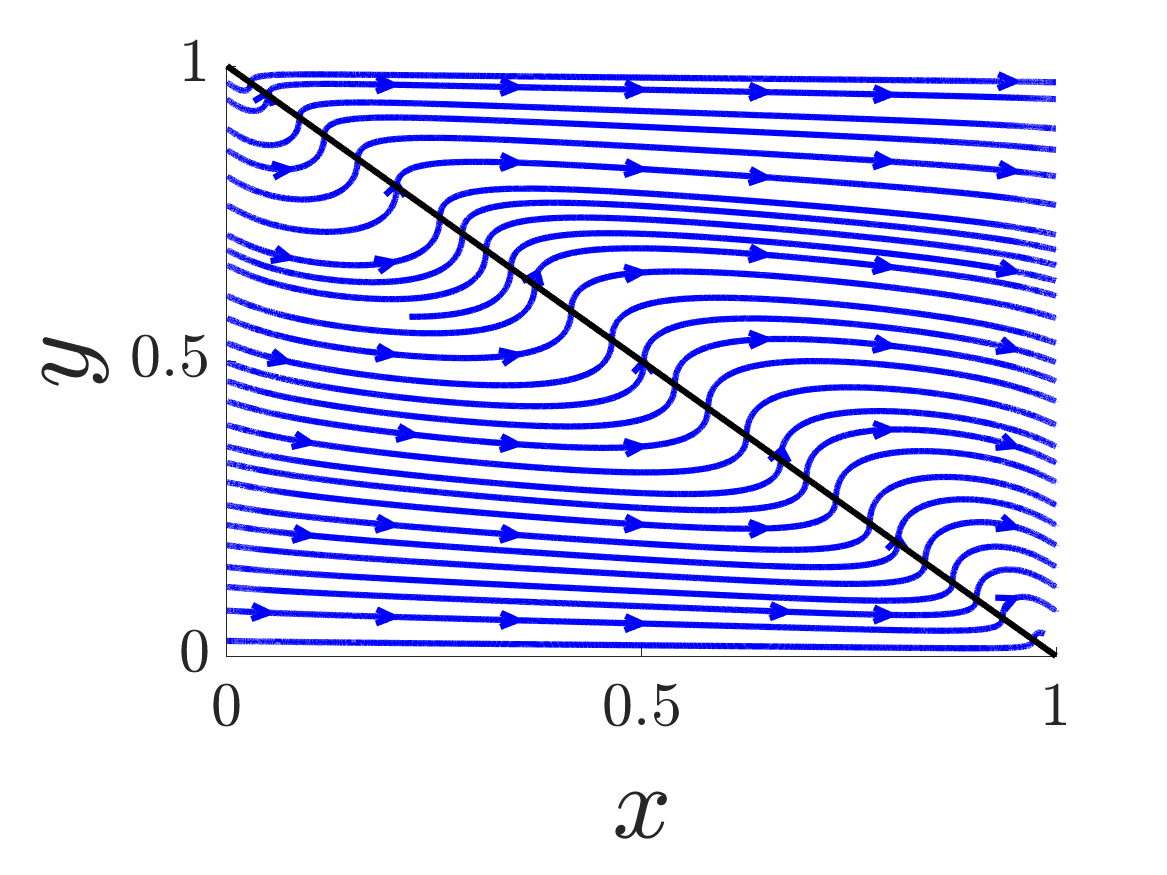}\\
    \vspace{-.5cm}(m) \hfill (n) \hfill (o) \hfill (p) \hfill \hfill \hfill \\ 
  \includegraphics[width=.24\linewidth]{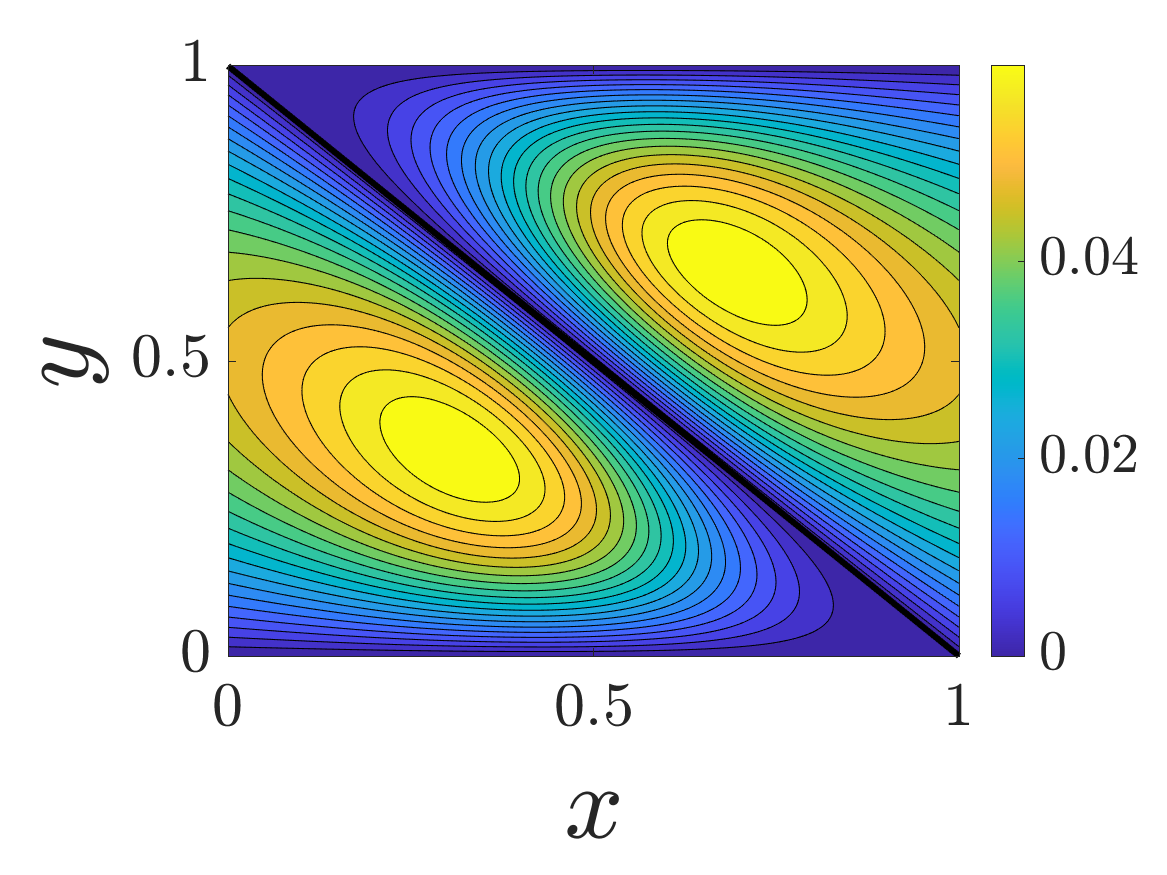}
  \includegraphics[width=.24\linewidth]{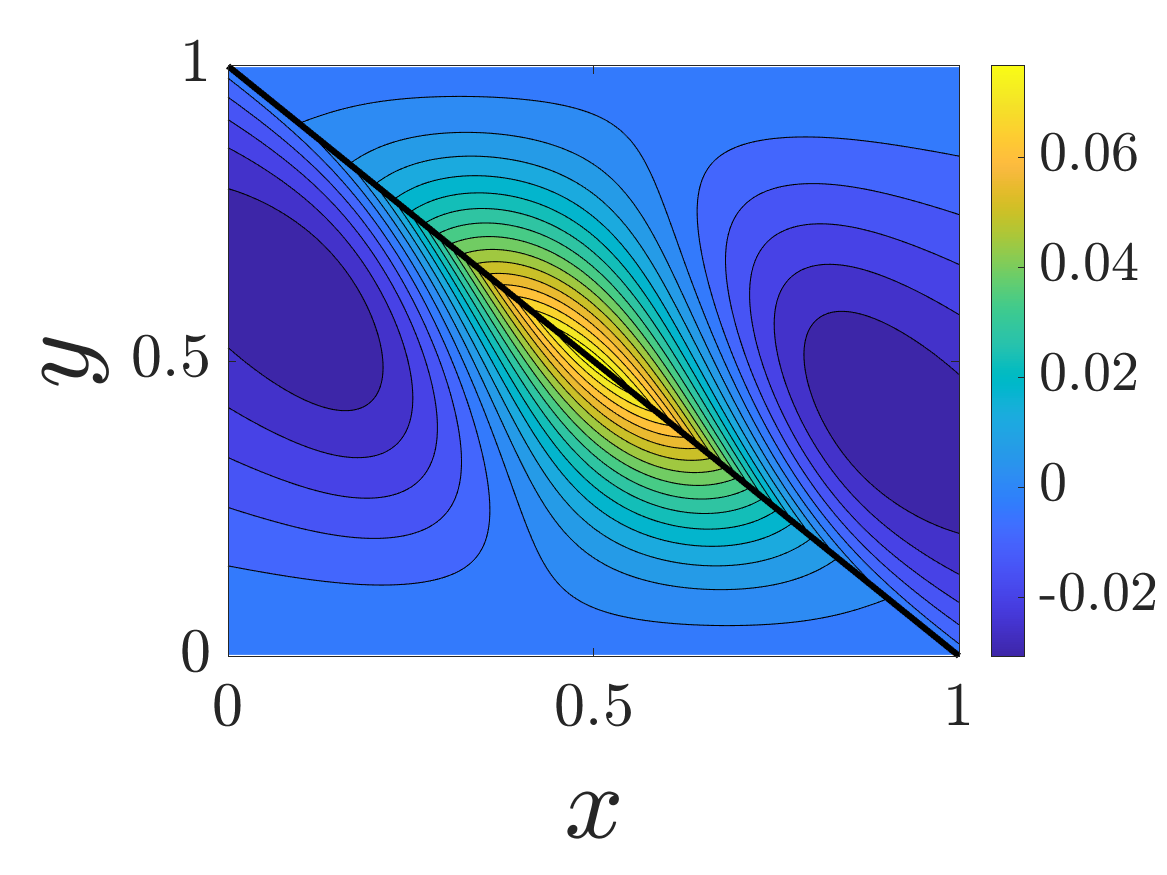}
  \includegraphics[width=.24\linewidth]{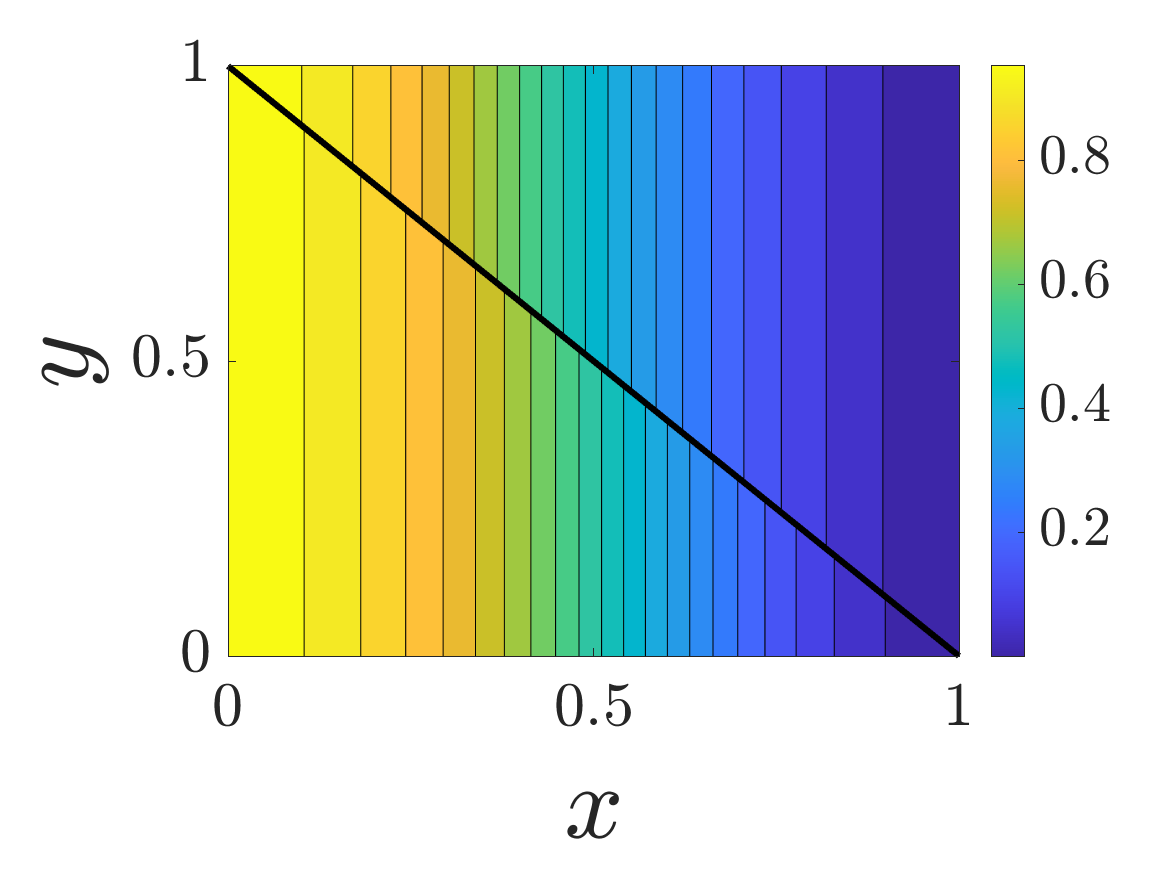}
  \includegraphics[width=.24\linewidth]{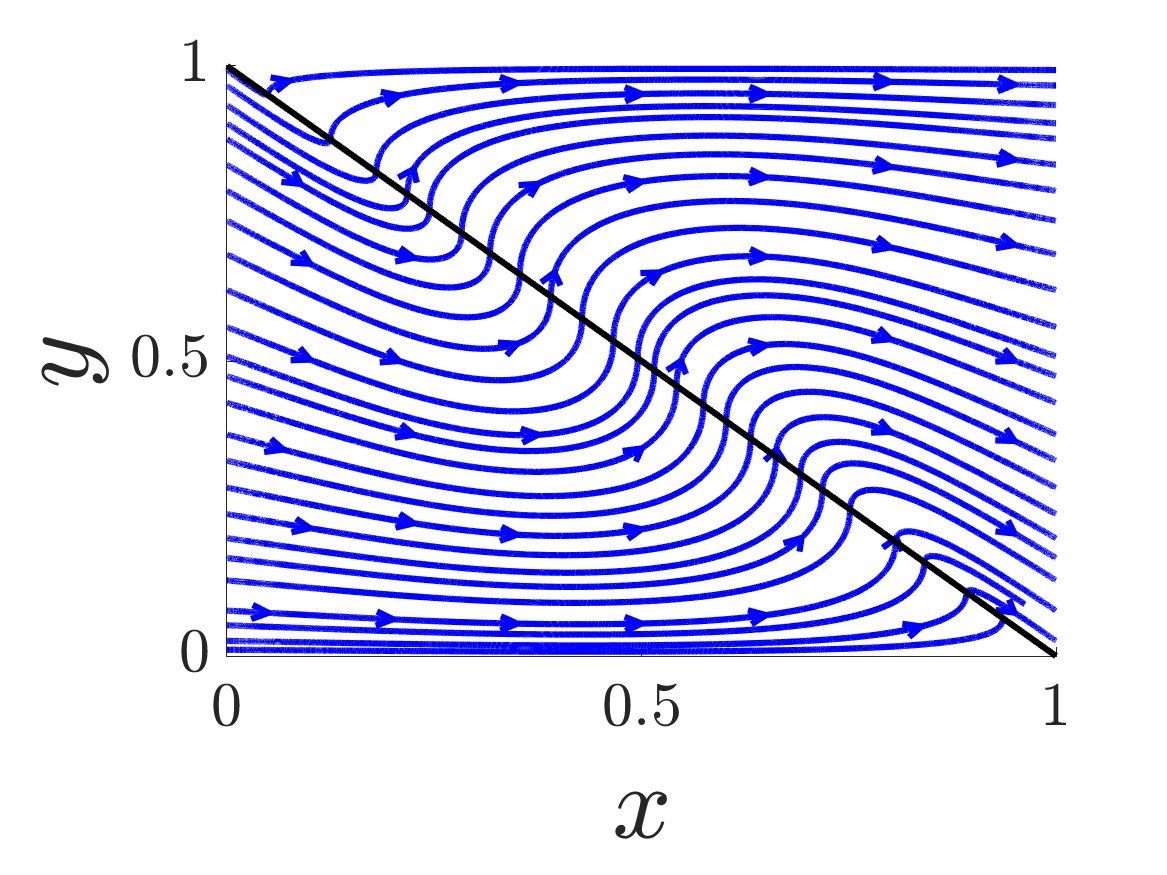}\\
  \caption{The streamwise velocity (a, e), normal velocity (b, f), pressure field (c, g) and streamlines (d, h) for (a--d) $k = 1\times10^{-7}$ and (e--h) $k = 1\times10^{-5}$ without separators using \eqref{eq:bvp_1}. The streamwise velocity (i,\,m), normal velocity (j,\,n), pressure field (k,\,o) and streamlines (l,\,p) for (i--l) $k = 1\times10^{-7}$ and (m--p) $k = 1\times10^{-5}$ without separators using \eqref{eq:bvp_3}.
  }
    \label{fig:flow}
\end{figure*}

In Fig.~\ref{fig:flow}, we examine how $k$ affects the streamwise velocity field ($u$), normal velocity field ($v$), pressure field ($p$) and streamlines $(u, v)$. 
We compare the flow field without separators (panels a--h) and with separators (panels i--p), where $u$ is either no slip or symmetric at $y = 0, 1$.
For high permeability ($k \gg 1$), shown in panels (e--h, m--p), the filter offers little resistance to the flow passing through it. 
As a result, the pressure fields in regions $D_1$ and $D_2$ equilibrate and vary linearly due to the total pressure drop across the filter. 
This pressure drop generates a streamwise velocity in $D_1$ and $D_2$, which is converted into a normal velocity through the fibre sheet. 

For low permeability ($k \ll 1$), shown in panels (a--d, i--l), the filter strongly resists normal flow, effectively blocking fluid from passing through the fibre sheet. 
Consequently, the V-shaped filter half-period experiences minimal through-flow, and the pressure fields in $D_1$ and $D_2$ are approximately uniform. 
The pressure in $D_1$ is equal to the inlet pressure, while the pressure in $D_2$ matches the outlet pressure. 
Due to the small pressure gradients in these regions, the streamwise velocity in $D_1$ and $D_2$ remains low, leading to only a small normal velocity through the fibre sheet. 
\update{These results illustrate two limiting behaviours of the long-wave model: a linear pressure gradient with a parabolic velocity profile at high permeability, and an almost uniform pressure field with weak through-flow at low permeability. 
The permeability relation $k = \hat{\mu}\hat{\kappa}/(\hat{t}\epsilon^3)$ (Section~\ref{sec:formulation}) can therefore be used to tune filter sheet properties, thickness, pleat geometry and fluid viscosity to maintain an even pressure drop, minimise local pollutant accumulation and balance filtration efficiency against energy consumption in V-shaped filter applications. 
}

\subsection{Potential pollutant removal rates}
\label{subsec:Potential pollutant removal rates}

\update{To illustrate the findings, we analyse results from the laminar flow experiments in V-shaped filters~\cite{del2002air, zhang2022operating, mrad2021local, al2011effect}, summarised in Table~\ref{tab:flow_rates}.}
\update{We perform a sensitivity analysis on model parameters to explore variability in airflow and pollutant removal performance.} 
\update{To estimate the airflow rate for a V-shaped filter, we scale the measured airflow per half-period, assuming an average filter width of 0.2 to 2 m and 100 to 300 repeating half-periods, consistent with typical ventilation systems~\cite{tomlinson2025harnessing}.} 
An increase in filter sheet permeability, driven by variations in fibre diameter, porosity or geometry, from the values reported by~\cite{zhang2022operating} to those reported by~\cite{mrad2021local}, results in a doubling of the airflow rate through the filter from approximately 1 to 2 m$^3$/s.
Higher permeability reduces flow resistance, allowing increased airflow, but decreases filtration efficiency from 99.97 to 52.2\%, likely due to a shorter residence time of particles within the porous structure of the filter medium. 
Despite the reduction in filtration efficiency, the increased airflow enhances the maximum potential atmospheric removal rate from $3.4 \times 10^{-7}$ to $3.5 \times 10^{-7}$ mol/s for PM$_{2.5}$, $1.0 \times 10^{-6}$ to $1.1 \times 10^{-6}$ mol/s for NO$_\text{x}$ and $9.2 \times 10^{-6}$ to $9.6 \times 10^{-6}$ mol/s for CH$_4$ ($7.7 \times 10^{-4}$ to $8.1 \times 10^{-4}$ mol/s for CO$_2$e, 20-year GWP), as a greater volume of air is processed per unit time.
\update{We calculate the potential atmospheric removal rate using~\eqref{eq:pollutant_removal_rate} and Tables~\ref{tab:scaling_paras}--\ref{tab:flow_rates}. 
These tables summarise the parameters used in these estimates, including removal probability $r$, removal efficiency $\eta$, ambient concentration $\hat{c}$ and flow rate $\hat{Q}$, together with literature sources defining lower- and upper-bound scenarios. 
}

\begin{table*}[t!]
    \centering
    \renewcommand{\arraystretch}{1.3} 
    \setlength{\tabcolsep}{4pt} 
    \scriptsize
    \update{
    \begin{tabular}{lcccccc}
        \toprule
        Parameter & Symbol & Unit & PM$_{2.5}$ & NO$_\text{x}$ & CH$_4$ \\
        Removal probability & $r$ & -- & [0.1, 1.0] & [0.05, 0.5] & [0.01, 0.1] \\
        Ambient concentration & $\hat{c}$ & mol/m$^3$ & [2.7$\times10^{-7}$, 3.3$\times10^{-7}$] & [1$\times10^{-6}$, 2$\times10^{-6}$] & [7$\times10^{-5}$, 9$\times10^{-5}$] \\
        \bottomrule
    \end{tabular}   
    }
     \caption{\update{Summary of parameters used to estimate the pollutant removal potential of V-shaped catalytic filters. 
    Removal probability $r$: PM$_{2.5}$~\cite{del2002air, mrad2021local, zhang2022operating, al2011effect}, NO$_\text{x}$~\cite{li2022recent, tomlinson2025harnessing} and CH$_4$~\cite{tsopelakou2024exploring, tomlinson2025harnessing}. 
    Ambient concentration $\hat{c}$: PM$_{2.5}$~\cite{mrad2021local, cams2023eac4}, NO$_\text{x}$~\cite{li2022recent, cams2023eac4} and CH$_4$~\cite{tsopelakou2024exploring, cams2023eac4}. 
    The ranges correspond to lower- and upper-bound estimates. 
    }}
    \label{tab:scaling_paras}
\end{table*}

\begin{table*}[t!]
    \centering
    \renewcommand{\arraystretch}{1.3} 
    \setlength{\tabcolsep}{4pt} 
    \scriptsize
    \update{
    \begin{tabular}{lccccc}
        \toprule
        Authors  & Flow rate ($\hat{Q}$) & Eff. ($\eta$) & Removal rate PM$_{2.5}$ ($\hat{q}$) &  Removal rate NO$_\text{x}$ ($\hat{q}$) & Removal rate CH$_4$ ($\hat{q}$) \\
        Fabbro \textit{et al.} \cite{del2002air} & 0.61 & 99.97 & [$1.7\times10^{-8}$,\,$2.0\times10^{-7}$] & [$3.1\times10^{-8}$,\,$6.1\times10^{-7}$] & [$4.3\times10^{-7}$,\,$5.5\times10^{-6}$] \\
        Mrad \textit{et al.} \cite{mrad2021local} & 2.03 & 52.5 & [$2.9\times10^{-8}$,\,$3.5\times10^{-7}$] & [$5.3\times10^{-8}$,\,$1.1\times10^{-6}$] & [$7.5\times10^{-7}$,\,$9.6\times10^{-6}$] \\
        Zhang \textit{et al.} \cite{zhang2022operating} & 1.02 & 99.97 & [$2.8\times10^{-8}$,\,$3.4\times10^{-7}$] & [$5.1\times10^{-8}$,\,$1.0\times10^{-6}$] & [$7.1\times10^{-7}$,\,$9.2\times10^{-6}$] \\
        Al-Attar \textit{et al.} \cite{al2011effect} & 2.30 & 94 & [$5.8 \times10^{-8}$,\,$7.1 \times10^{-7}$] & [$1.1\times10^{-7}$,\,$2.2\times10^{-6}$] & [$1.5\times10^{-6}$,\,$2.0\times10^{-5}$] \\
        \bottomrule
    \end{tabular}  
    }
    \caption{\update{Examples of potential removal rates \eqref{eq:pollutant_removal_rate} of PM and gases using the laminar datasets~\cite{del2002air, mrad2021local, zhang2022operating, al2011effect} on V-shaped filters: their width ($\hat{W}\in[0.2,\,2]$ m), number of half-periods ($N\in[100,\,300]$), flow rate ($\hat{Q}$, m$^{3}$/s), filtration efficiency ($\eta$), potential removal rate of PM$_{2.5}$, NO$_\text{x}$ and CH$_4$ ($\hat{q}$, mol/s). 
    }}
    \label{tab:flow_rates}
\end{table*}

\update{Globally, we assume approximately $5\times10^8$ to $2\times10^{9}$ filters are in operation~\cite{tomlinson2025harnessing}, corresponding to a total flow rate of $0.3$ to $4.6$ Gm$^3$/s, based on Table 3.} 
\update{If catalyst technologies were integrated into these systems, the projected atmospheric removal rate could range from
$2.7\times10^{-5}$ to $4.5\times10^{-3}$ GtPM$_\text{2.5}$,
$2.3\times10^{-5}$ to $6.4\times10^{-3}$ GtNO$_\text{x}$,  $1.1\times10^{-4}$ to $2.0\times10^{-2}$ GtCH$_\text{4}$ per year ($9.2\times10^{-3}$ to $1.6\times10^{0}$ GtCO$_\text{2}$e per year, 20-year GWP for CH$_4$).
The estimated cost is \$$3.4\times10^{3}$ to $3.8\times10^{8}$ per tNO$_\text{x}$ and \$$1.1\times10^{3}$ to $7.8\times10^{7}$ per tCH$_4$ (\$$1.3\times10^{1}$ to $9.3\times10^{5}$ per tCO$_2$e) removed, assuming \$15 to \$43 per m$^2$ for coating the filter sheet~\cite{randall2024cost, hickey2024economics}.} 
\update{Variability in removal rates and costs is mainly due to engineering parameters such as flow rate, removal probability and catalyst efficiency, rather than ambient concentrations, so improved filter design could narrow these ranges.}
\update{These findings indicate that optimising permeability, together with catalytic integration, could allow filters to operate as large-scale atmospheric removal devices, removing more than 1 GtCO$_2$e per year as an upper bound at less than \$100 per tCO$_2$e as an lower bound~\cite{ipcc2023synthesis}.} 

\section{Discussion} \label{sec:discussion}

This study has used a long-wave theory to model laminar flow through V-shaped filters for pollutant removal. 
The long-wave model simplifies the airflow problem in the V-shaped filter, preserving physical details and improving computational efficiency compared to the complete problem. 
Our results agree with numerical simulations and experimental data, confirming the long-wave model's applicability to diverse V-shaped filter configurations.
The long-wave model captures key relationships between pressure drop, permeability, velocities and flow resistance, predicting the removal efficiency and permeability based on measurable filter properties such as fibre diameter, porosity, filter length, half-height and thickness. 

Our findings explain the critical trade-off in V-shaped filter design, underscoring the need for an optimised balance between flow rate, pollutant removal and energy efficiency.
Increasing fibre diameter and porosity enhances airflow but reduces available surface area for particle capture, lowering removal efficiency.
Conversely, increasing the aspect ratio and thickness of the filter improves pollutant capture but increases flow resistance, raising the energy demands of the ventilation system.
\update{The cost estimates in Section~\ref{subsec:Potential pollutant removal rates} reflect a range of scenarios based on realistic parameters under optimistic laboratory conditions. 
Actual economic feasibility will depend strongly on local factors such as catalyst lifetime, energy input, airflow rate and filter renewal frequency. 
These aspects remain important topics for future work. 
}
By modifying filter surfaces with other removal technologies, these systems could capture and convert CO$_2$, CH$_4$, N$_2$O, NO$_\text{x}$ and VOCs by integrating carbon capture, catalysis and air purification within existing infrastructure.
These multifunctional filters would enhance both indoor and outdoor air quality, particularly in environments with stringent air quality regulations.
Future work could explore adaptive filtration strategies that incorporate real-time monitoring and dynamic adjustments to pollutant concentrations, leading to self-regulating and more efficient air purification.

Nevertheless, the limitations of the approach must be acknowledged. 
The long-wave model’s assumption of steady-state, laminar flow excludes non-linear effects such as clogging and turbulence, which may alter performance over extended operation. 
While regular filter maintenance mitigates some clogging, real-world conditions may introduce unpredictable accumulation patterns or high dust loads.
\update{The long-wave model is valid for V-shaped filters in ventilation systems with approximately ${Re} < 2000$, defined using the velocity and height of the pleat, corresponding to laminar flow. 
Transitional behaviour typically occurs around ${Re} \approx 2000$ to $3000$, where inertial and emerging turbulent effects modify the pressure drop, mixing and pollutant transport~\cite{mcpherson2009subsurface}. 
These effects are partially reflected in datasets with non-linear pressure--flow relationships (Fig.~\ref{fig:all}). 
For approximately $Re > 3000$, turbulence dominates the flow. 
Consequently, predictions of pressure drop and pollutant transport using the long-wave model may be underestimated for some applications. 
Future work incorporating turbulent flow effects via direct numerical simulations will allow a comparison of laminar and turbulent performance  (Table~\ref{tab:experiments})~\cite{rebai2010semi, tronville2003minimization, theron2017numerical}. 
}
With these extensions, the model could potentially be generalised to flue gas treatment and automotive exhaust systems~\cite{li2022recent, van2001science}.
While the exponential and Kozeny--Carman relations provide reliable estimates of removal efficiency and permeability, their empirical nature may require further calibration for different types of filters (e.g., U-shaped) or operating conditions (e.g., turbulence). 
\update{Experimental campaigns are currently underway to validate removal efficiencies for CH$_4$, NO$_\text{x}$ and VOCs in V-shaped catalytic filters under realistic ventilation conditions, as no such studies currently exist. 
Results from Malayeri \textit{et al.}~\cite{malayeri2019modeling}, reviewing photo-catalytic VOC degradation in airflow through monoliths and packed beds; Tsopelakou \textit{et al.} (in preparation), demonstrating photo-catalytic CH$_4$ conversion on filter surfaces under laboratory conditions; Li \textit{et al.}~\cite{li2022recent}, reviewing catalytic filters for integrated dust and NO$_\text{x}$ removal from flue gas; and Songxuan \textit{et al.}~\cite{songxuan2025application}, demonstrating pilot-scale denitrification, dioxin breakdown and PM capture on catalytic filter substrates, collectively support the feasibility of the integrated pollutant removal concept presented here. 
}

\section{Conclusions}

\update{This study developed and validated a non-linear long-wave model for laminar flow through V-shaped filters (with and without separators), integrating catalytic pollutant-removal. 
The model reproduces experimental pressure--flow relationships and quantifies trade-offs between permeability, energy efficiency and pollutant removal. 
Limitations include the assumption of laminar flow and uncertainty in catalytic performance under real-world conditions. 
Future work will extend the model to transitional and turbulent regimes and validate catalytic removal experimentally. 
Overall, the framework unites theoretical, numerical and experimental analyses of laminar flow in V-shaped filters, providing a blueprint for optimising catalytic filtration systems capable of mitigating PM, GHGs and VOCs. 
    }

\section*{CRediT authorship contribution statement}

\textbf{Samuel D. Tomlinson, Aliki M. Tsopelakou, Tzia Ming Onn, Steven R. H. Barrett, Adam M. Boies, Shaun Fitzgerald}: Conceptualization, Methodology, Investigation, Writing.

\section*{Declaration of competing interest}

The authors declare that they have no known competing financial interests or personal relationships that could have appeared to influence the work reported in this paper.

\section*{Acknowledgments}

We acknowledge Grantham Foundation for supporting this research.

\section*{Data availability}

Data will be made available on request.

\appendix

\section{Asymptotic solutions} \label{app:asym}

\subsection{Large-permeability solution}

When the permeability of the filter sheet is large, $\kappa \gg 1$, we can simplify the system of ODEs for pressure without separators \eqref{eq:bvp_1}--\eqref{eq:bc_2}. 
In Fig.~\ref{fig:flow}(a--d), the filter offers little resistance to the flow so that the pressure field becomes uniform across the channel. 
We expand the pressure field as follows
\begin{equation} \label{eq:pressure_expansion}
    p_0^- = p_{00}^- + p_{01}^-/\kappa + ..., \quad p_0^+ = p_{00}^+ + p_{01}^+/\kappa + ...,
\end{equation}
and substitute \eqref{eq:pressure_expansion} into \eqref{eq:bvp_1}--\eqref{eq:bc_2}.
At leading order, along $y = s(x)$, we have $p^-_{00} = p^+_{00} = p_{00}$, such that
\begin{subequations} \label{eq:bvp_1_0}
\begin{align}
   p_{00xx} s^3 /3 - p_{00x} s^2 s_x &= \kappa (p^-_{01} - p^+_{01}), \\
   p_{00xx}(s-1)^2/3 - p_{00x} (s-1)^2 s_x &= \kappa (p^-_{01} - p^+_{01}).
\end{align}
\end{subequations}
At $x = 0$, the inlet and no-gradient condition is 
\begin{equation} \label{eq:bc_1_0}
    p_{00} = 1, \quad p_{00x} = 0,
\end{equation}
and at $x = 1$, the outlet and no-gradient condition is 
\begin{equation} \label{eq:bc_2_0}
    p_{00} = 0, \quad p_{00x} = 0.
\end{equation}
Subtracting the equations in \eqref{eq:bvp_1_0}, \eqref{eq:bvp_1_0}--\eqref{eq:bc_2_0} reduces to 
\begin{equation} \label{eq:bvp_1_combined}
   p_{00xx} (s^3 - (s-1)^3) /3 - p_{00x} (-s^2 + (s-1)^2)s_x = 0, \quad p_{00}(0) = 1, \quad p_{00}(1) = 0,
\end{equation}
such that the leading-order pressure distribution is
\begin{equation} \label{eq:pres_sol}
    p_{00} = \frac{\pi - 3\arctan(\sqrt{3}(2x-1))}{2\pi}.
\end{equation}
Hence, the leading-order flow rate, evaluated by substituting \eqref{eq:pres_sol} into \eqref{eq:flux_1}, is given by
\begin{equation} \label{eq:flux_1a}
    Q = \frac{W \sqrt{3}}{4 \pi} + ....
\end{equation}
We can perform the same procedure for the system of ODEs for pressure with separators \eqref{eq:bvp_3} to recover \eqref{eq:pres_sol}.
Hence, the leading-order flow rate, evaluated by substituting \eqref{eq:pres_sol} into \eqref{eq:flux_2}, is given by
\begin{equation} \label{eq:flux_2b}
    Q = \frac{W \sqrt{3}}{16 \pi} + ....
\end{equation}
\eqref{eq:flux_1a}--\eqref{eq:flux_2b} agree with the results shown in Fig.~\ref{fig:flux}.

\subsection{Small-permeability solution}

When permeability is small, $\kappa \ll 1$, we can simplify the system of ODEs for pressure without separators. 
In Fig.~\ref{fig:flow}(e--h), the filter blocks the flow such that the pressure field becomes constant and equal to the inlet condition in $D_1$ and equal to the outlet condition in $D_2$. 
We expand the pressure field as follows
\begin{equation} \label{eq:pressure_expansion_2}
    p_0^- = p_{00}^- + \kappa p_{01}^- + ..., \quad p_0^+ = p_{00}^+ + \kappa p_{01}^+ + ...,
\end{equation}
and substitute \eqref{eq:pressure_expansion_2} into \eqref{eq:bvp_1}--\eqref{eq:bc_2}.
At leading order, along $y = s(x)$, 
\begin{subequations} \label{eq:bvp_1_0_2}
\begin{align}
    p_{00xx}^- s^3 /3 - p_{00x}^- s^2 s_x &= 0, \\
   p_{00xx}^+(s-1)^2/3 - p^+_{00x} (s-1)^2 s_x &= 0.
\end{align}
\end{subequations}
At $x = 0$, the inlet and no-gradient condition is 
\begin{equation} \label{eq:bc_1_0_2}
    p_{00}^- = 1, \quad p_{00x}^- = 0,
\end{equation}
and at $x = 1$, the outlet and no-gradient condition is 
\begin{equation} \label{eq:bc_2_0_2}
    p_{00}^+ = 0, \quad p_{00x}^+ = 0.
\end{equation}
The leading-order pressure distribution in $D_1$ and $D_2$ is 
\begin{equation} \label{eq:pres_sol_2}
    p_{00}^- = 1, \quad p_{00}^+ = 0.
\end{equation}
Hence, the leading-order flow rate, evaluated by substituting \eqref{eq:pres_sol_2} into \eqref{eq:flux_1}, is given by
\begin{equation} \label{eq:lo_flux_2}
    Q = O(\kappa),
\end{equation}
which agrees with the numerical results shown in Fig.~\ref{fig:flux}.
We can perform the same procedure for the system of ODEs for pressure with separators \eqref{eq:bvp_3} to recover \eqref{eq:pres_sol_2} and \eqref{eq:lo_flux_2}.

\section{Comparison between Darcy and Darcy--Forchheimer models}
\label{app:comp}

\begin{figure}
    \centering
    \hfill\includegraphics[width=0.49\linewidth,trim={0 0 1cm 0},clip]{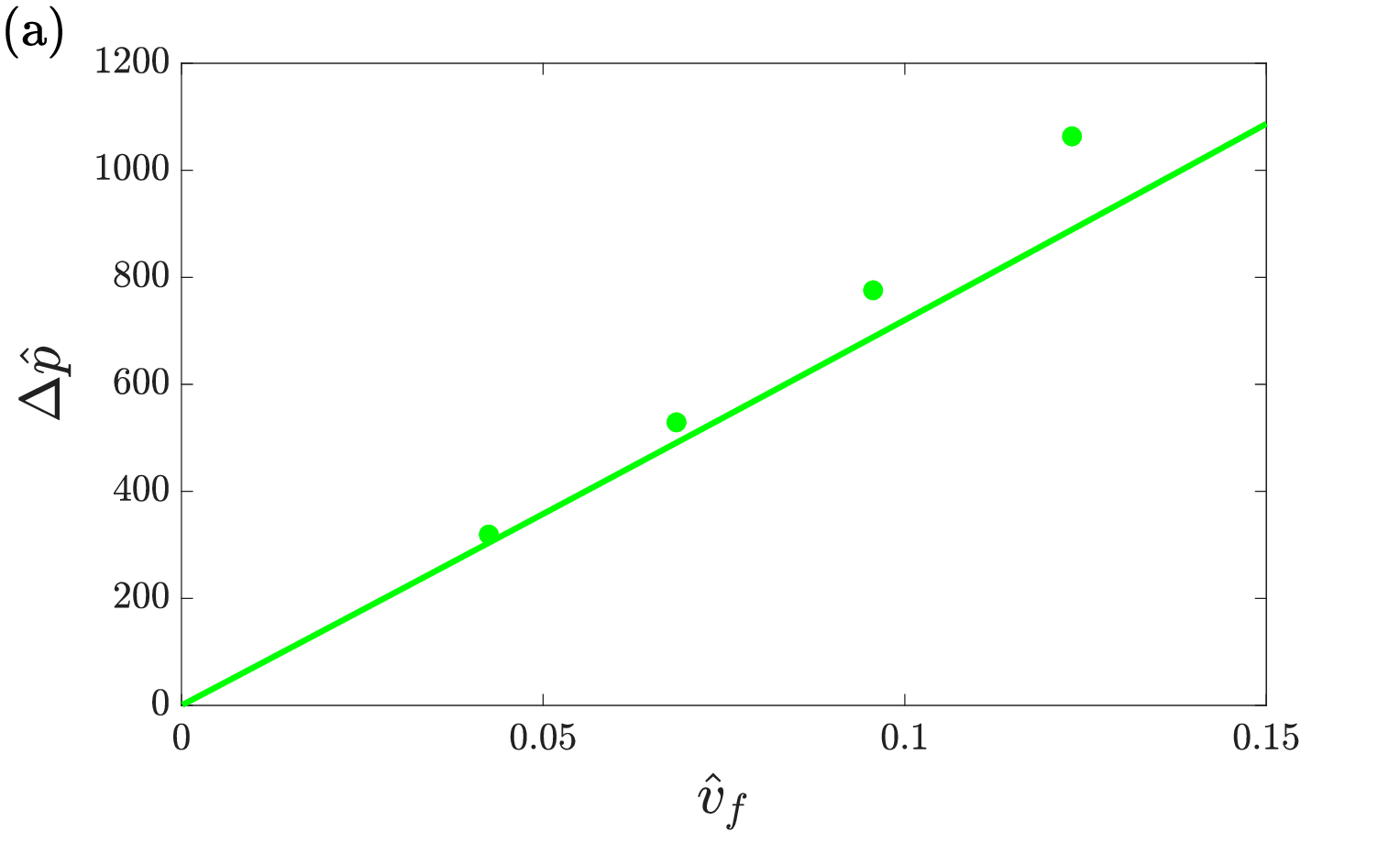} \hfill \includegraphics[width=0.49\linewidth,trim={0 0 1cm 0},clip]{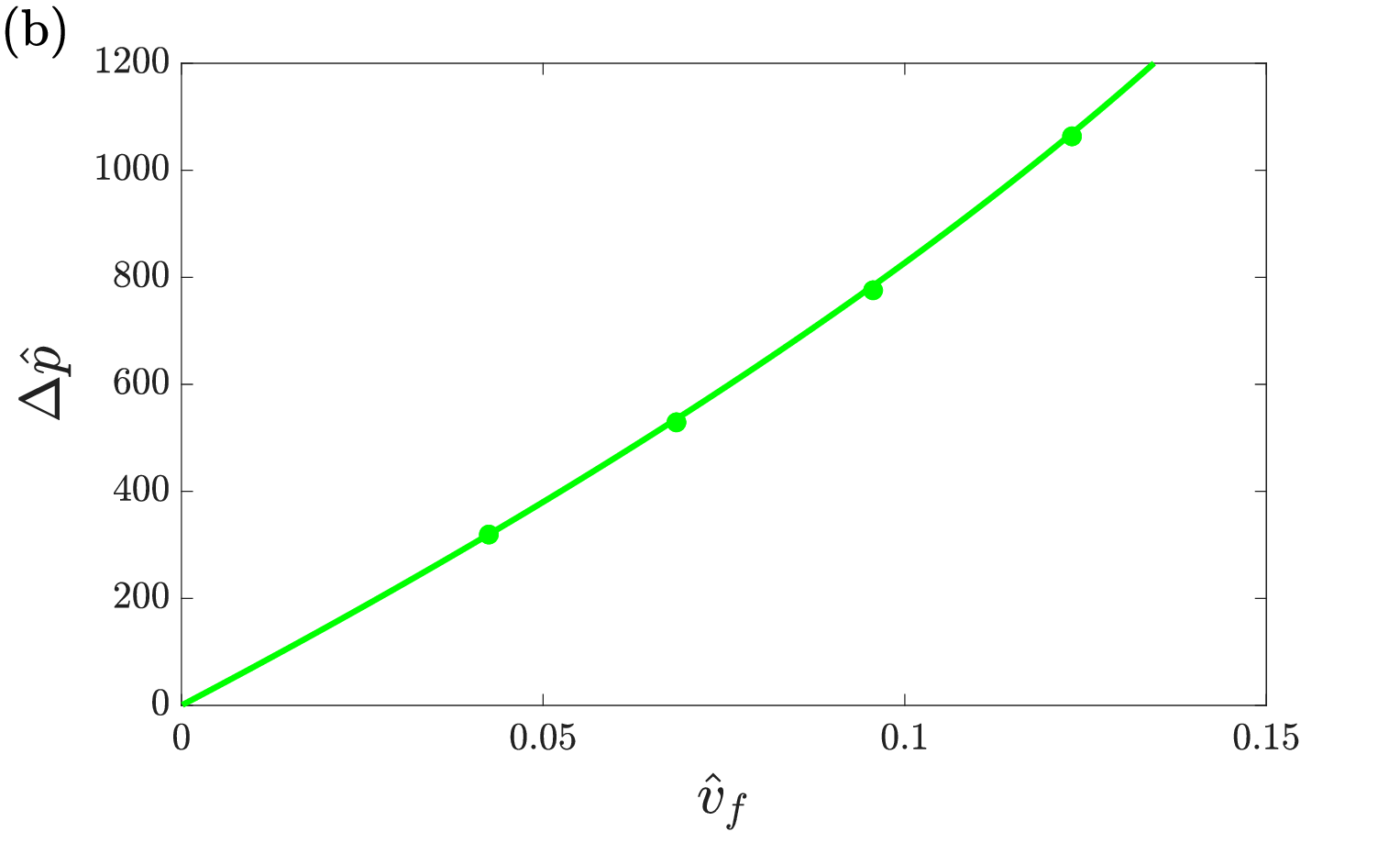}\hfill\hfill\hfill\\
    \caption{Comparison of the pressure drop ($\Delta \hat{p}$) with the filtration velocity ($\hat{v}_f$) predicted using the long-wave model \eqref{eq:bvp_1} (solid lines), for (a) $\hat{\beta} = 0$ s/m and (b) $\hat{\beta} = 1.4\times10^{-4}$ s/m, with experiments (symbols) in~\cite{del2002air} for $\epsilon = 0.037$.}
    \label{fig:app_fig}
\end{figure}

\update{Fig.~\ref{fig:app_fig} compares the Darcy and Darcy--Forchheimer models for a representative filter from Fabbro \textit{et al.}~\cite{del2002air}, showing an absolute error of $O(10)$ Pa at $\hat{v}_f = 0.0375$ m/s (where $Re = 124$ and nonlinear effects are weaker, panel a--b), $O(100)$ Pa for $\hat{v}_f = 0.0878$ m/s with $\hat{\beta} = 0$ s/m (panel a) and $O(10)$ Pa for $\hat{v}_f = 0.0878$ m/s with $\hat{\beta} = 1.4\times10^{-4}$ s/m (where $Re = 290$ and nonlinear effects are stronger, panel b), demonstrating the improved quantitative agreement when including inertial (Forchheimer) contributions.} 

\bibliographystyle{elsarticle-num}
\bibliography{refs}

\end{document}